\documentclass[preprint2]{proto}
\usepackage{natbib}
\bibpunct{(}{)}{;}{a}{,}{,}
\usepackage{times}
\usepackage{amsmath}
\usepackage{relsize}
\usepackage{color}
\usepackage[draft]{hyperref}

\graphicspath{{fig/}}

\newcommand{\sigs}{\sigma_s}
\newcommand{\sfrff}{\mathrm{SFR}_\mathrm{ff}}
\newcommand{\sfe}{\mathrm{SFE}}
\newcommand{\alphavir}{\alpha_\mathrm{vir}}
\newcommand{\mach}{\mathcal{M}_\mathrm{s}}
\newcommand{\macha}{\mathcal{M}_\mathrm{A}}
\newcommand{\phit}{\phi_t}
\newcommand{\phix}{\phi_x}
\newcommand{\tff}{t_\mathrm{ff}}
\newcommand{\ycut}{y_\mathrm{cut}}
\newcommand{\chisqred}{\chi^2_\mathrm{red}}
\newcommand{\sigsfr}{\Sigma_\mathrm{SFR}}
\newcommand{\siggas}{\Sigma_\mathrm{gas}}
\newcommand{\cs}{c_\mathrm{s}}
\newcommand{\va}{v_\mathrm{A}}
\newcommand{\meanrho}{\rho_0}
\newcommand{\scrit}{s_\mathrm{crit}}
\newcommand{\deriv}{\,\mathrm{d}}
\newcommand{\rhocrit}{\rho_\mathrm{crit}}
\newcommand{\sigsf}{\Sigma_\mathrm{SF}}
\newcommand{\tsf}{t_\mathrm{SF}}
\newcommand{\msol}{\mbox{$M_{\sun}$}}
\newcommand{\pc}{\mathrm{pc}}
\newcommand{\yr}{\mathrm{yr}}

\newcommand{\ee}[1]{\mbox{${} \times 10^{#1}$}}
\newcommand{\mean}[1]{\mbox{$\langle#1\rangle$}} 
\newcommand{\lsun}{\mbox{L$_\odot$}}
\newcommand{\msun}{\mbox{M$_\odot$}}
\newcommand{\texcess}{\mbox{$t_{excess}$}}
\newcommand{\mstar}{\mbox{$M_{\star}$}}
\newcommand{\sfr }{\mbox{$\dot M_{*}$}}
\newcommand{\sigmasfr}{\mbox{$\Sigma_{\rm SFR}$}}
\newcommand{\sigmagas}{\mbox{$\Sigma_{gas}$}}
\newcommand{\msunpc}{\mbox{M$_\odot$ pc$^{-2}$}}
\newcommand{\eps}{\mbox{$\epsilon$}}
\newcommand{\sigmamol}{\mbox{$\Sigma_{mol}$}}
\newcommand{\mgas}{\mbox{$M_{gas}$}}
\newcommand{\tdep}{\mbox{$t_{dep}$}} 
\newcommand{\cmv}{\mbox{cm$^{-3}$}}
\newcommand\av{\mbox{$A_V$}}
\newcommand{\coo}{\mbox{$^{13}$CO}}

\newcommand{\beq}   {\begin{equation}}
\newcommand{\eeq}   {\end{equation}}
\def\symbol#1{\ifmmode#1\else$#1$\fi}
\def\calm{\symbol{{\cal M}}}

\def\aviro{\alpha_{\rm vir,\,O}}
\def\avirt{\alpha_{\rm vir,\,T}}

\def\avircl{\alpha_{\rm cl}}
\def\cl{{\rm cl}}

\def\cs{c_{\rm s}}

\def\sod{\sigma_{\rm v,1D}}

\begin{document}

\title{\textbf{\LARGE The Star Formation Rate of Molecular Clouds}}

\author {\textbf{\large Paolo Padoan}}
\affil{\small\em University of Barcelona}
\author {\textbf{\large Christoph Federrath}}
\affil{\small\em Monash University}
\author {\textbf{\large Gilles Chabrier}}
\affil{\small\em Ecole Normale SupŽrieure de Lyon}
\author {\textbf{\large Neal J. Evans II}}
\affil{\small\em The University of Texas at Austin}
\author {\textbf{\large Doug Johnstone}}
\affil{\small\em University of Victoria }
\author {\textbf{\large Jes K. J{\o}rgensen}}
\affil{\small\em University of Copenhagen}
\author {\textbf{\large Christopher F. McKee}}
\affil{\small\em University of California, Berkeley}
\author {\textbf{\large {\AA}ke Nordlund}}
\affil{\small\em University of Copenhagen}

\begin{abstract}
\baselineskip = 11pt
\leftskip = 0.65in
\rightskip = 0.65in
\parindent=1pc
{\small We review recent advances in the analytical and numerical modeling of the star formation rate in molecular clouds
and discuss the available observational constraints. We focus on molecular clouds as the fundamental star formation
sites, rather than on the larger-scale processes that form the clouds and set their properties. Molecular clouds are shaped
into a complex filamentary structure by supersonic turbulence, with only a small fraction of the cloud mass channeled into
collapsing protostars over a free-fall time of the system. In recent years, the physics of supersonic turbulence has
been widely explored with computer simulations, leading to statistical models of this fragmentation process, and to
the prediction of the star formation rate as a function of fundamental physical parameters of molecular clouds, such as
the virial parameter, the rms Mach number, the compressive fraction of the turbulence driver, and the ratio of gas
to magnetic pressure. Infrared space telescopes, as well as ground-based observatories have provided unprecedented
probes of the filamentary structure of molecular clouds and the location of forming stars within them. \\~\\~\\~}

\end{abstract}

\section{\textbf{INTRODUCTION}}

Understanding and modeling the star formation rate (SFR) is a central goal of a theory of star formation. Cosmological
simulations of galaxy formation demonstrate the impact of SFR models on galaxy evolution \citep{2013ApJ...770...25A},
but they neither include star formation self-consistently, nor its feedback mechanisms.
They must rely instead on suitable subgrid-scale models to include the effects of star formation and feedback.

The SFR has been a fundamental problem in astrophysics since the 1970's, when it was shown that the gas depletion 
time in our Galaxy is much longer than any characteristic free-fall time
of star forming gas \citep{1974ARA&A..12..279Z,1997ApJ...476..166W}. The same problem applies to individual MCs, where the
gas depletion time is much longer than the fee-fall time as well \citep[e.g.][]{2009ApJS..181..321E}.
It has been argued that for local clouds \citep{2007ApJ...654..304K}, as well as for disk and starburst galaxies at low and high 
redshift \citep{2012ApJ...745...69K,2013arXiv1307.1467F}, the gas depletion time is always of the order of 100 free-fall times.

The first solution to the SFR problem was proposed under the assumption that MCs are supported against gravity by relatively strong
magnetic fields, with star formation resulting from the contraction of otherwise subcritical cores by ambipolar drift \citep{1987ARA&A..25...23S}. 
The relevant timescale for star formation would then be the ambipolar drift time, much longer than the free-fall time in MCs. By accounting for
the effect of photoionization on the coupling of gas and magnetic field, \citet{1989ApJ...345..782M} derived a model for the SFR that predicted gas
depletion times consistent with the observations.

When it became possible to carry out relatively large three-dimensional simulations of magneto-hydrodynamic (MHD) turbulence  in the 1990s,
the focus of most SFR studies gradually moved from the role of magnetic fields to the role of supersonic turbulence, as described in a number of reviews 
\citep{2004ARA&A..42..275S,2004ARA&A..42..211E,2004RvMP...76..125M,2007prpl.conf...63B,2007ARA&A..45..565M}. \citet{1999ApJ...526..279P}
showed that the Zeeman-splitting measurements of the magnetic field in MCs were likely to indicate a mean magnetic field
strength much weaker than previously assumed \citep[see also][]{2008ApJ...686L..91L,2009ApJ...702L..37L,2012MNRAS.420.3163B}. Based on a body of
recent observational results, the review by \citet{2007ARA&A..45..565M} concludes that MCs are mostly supercritical instead of subcritical, in which case the
ambipolar-drift time cannot be the solution to the SFR problem.

Current simulations of star formation by supersonic turbulence show a high sensitivity
of the SFR to the virial parameter, defined as twice the ratio of turbulent kinetic energy to gravitational energy \citep{2011ApJ...730...40P,2012ApJ...761..156F,2012ApJ...759L..27P}, and thus to the ratio
between the free-fall time and the turbulence crossing time, as suggested by analytical models based on the turbulent fragmentation
paradigm \citep{2005ApJ...630..250K,2011ApJ...730...40P,2011ApJ...743L..29H,2012ApJ...761..156F,2013ApJ...770..150H}.
Interestingly, the models and simulations also show that the magnetic field is still needed to make the SFR as low as observed in MCs,
even if the SFR is not controlled by ambipolar drift.

Most studies of the SFR published after Protostars and Planets V have focused on the role of turbulence, so this review revolves around
supersonic turbulence as well. Besides the theoretical ideas and the numerical simulations, we discuss the observations that are used to
constrain the SFR in MCs and have a potential to test the theory. Extragalactic observations can be used to constrain SFR models as well \citep{2012ApJ...745...69K,2013arXiv1307.1467F}.
They present the advantage of a greater variety of star formation environments relative to local MCs, but they are not as detailed as studies of
nearby clouds. Although the extragalactic field is rapidly evolving and very promising, the observational side of this review is limited to studies
of local clouds, and we only briefly discuss the issue of relating the local studies to the extragalactic literature. For more information on this topic,
we refer the reader to the recent review by \citet{2012ARA&A..50..531K}, and to the chapter by Dobbs et al. in this book.

We start by reviewing observational estimates of the SFR in MCs in Section 2, with a critical discussion of the methods
and their uncertainties. We also contrast the methods based on the direct census of stellar
populations in MCs, with indirect ones used to determine the SFR on larger scales in extragalactic studies.

We then review the theoretical models in Section 3, focusing on those that account for the turbulent nature of the
interstellar medium, and discuss their differences, limitations, and potential for further development.
The theory relies on statistical results from numerical studies of supersonic magnetohydrodynamic (MHD) turbulence, particularly
on the probability distribution of gas density, whose dependence on fundamental physical parameters
has been recently quantified in several studies. Besides the distribution of gas density, the key ingredient
in the theory is the concept of a critical density for star formation. However, MCs are characterized by
highly non-linear turbulent motions, producing shocks and filaments that are often described as fractal
structures, casting doubt on such a threshold density concept. We discuss how the various models justify
this approximation, and how the observations support the idea of a critical density.

Turbulence simulations have been used to directly derive the SFR, by introducing self-gravity and sink
particles to trace gravitationally unstable and star-forming gas \citep[e.g.,][]{2010ApJ...713..269F}. Over the last two years, these simulations have been employed in vast parameter studies that can
be used to test and constrain theoretical models. In Section 4, we offer a critical view of numerical methods and experiments and we summarize the most important findings from the comparison between
simulations and theories.

In Section 5 we compare both theory and simulations to the observational estimates of the SFR. We draw
conclusions and outline future directions for both models and
observations in Section 6.

\section{\textbf{OBSERVATIONS}}

Over the last decade infrared studies from ground-based telescopes,
 such as 2MASS
\citep{1992ASPC...34..203K}, 
and from space with the Spitzer Space Telescope 
\citep{2004ApJS..154....1W}
and Herschel Space Observatory 
\citep{2010A&A...518L...1P}
have provided extended
surveys of the populations of young stellar objects and their
evolutionary stages, as well as their distributions within their parental
clouds. Together with large-scale extinction and submillimeter
continuum maps, which provide measurements of the cloud mass and column density,  
these observations allow direct estimates of the star formation
rates and efficiencies in different cloud regions.

In writing this chapter, the authors realized that many misunderstandings
were generated because of different conceptions of what a molecular cloud
is. The obvious case is the ``cloud in the computer" versus the ``cloud
imaged by observers", but even the latter is subject to definitional issues
because of evolving sensitivity and observing techniques. We thus
begin the next subsection with a discussion of techniques for
measuring clouds and conclude with the evolving
observational definition of a molecular cloud.

\bigskip
\noindent
\textbf{2.1
Measuring Cloud Mass and Surface Density: Methods and Uncertainties}
\bigskip

The distribution of mass in nearby molecular clouds has been determined
primarily from extinction  mapping in the visible and infrared, 
using Spitzer and 2MASS data,
and from ground-based submillimeter surveys of dust continuum emission. 
These observations allow the surface densities and masses 
to be derived without assumptions about abundance and excitation of gas tracers, 
such as CO and its rarer isotopologues. The estimation of the total cloud mass does, 
however, rely on a consistency in the dust to gas ratio within and across clouds.
These techniques have thus largely
replaced maps of CO for nearby clouds, although CO is still required to uncover 
cloud kinematics. For more distant clouds, maps of 
CO and other species are still used, and we discuss the issues arising for cloud
definition and properties in Section 2.3.

The advantages of the extinction mapping technique are twofold. 
First, the availability and sensitivity of large-area optical and infrared 
detectors allow large regions to be observed efficiently. 
Also, only the extinction properties of the intervening dust are required 
to convert from extinction to column density.
The major disadvantages are the lack of resolution available, unless the infrared
observations are extremely deep and the inability to measure very high 
column densities where the optical depth in the infrared becomes too large
to see background stars. 

Extinction maps from optical data have provided valuable measures of cloud
extents, masses, and surface densities for modest extinctions
(e.g., \citealt{1999A&A...345..965C,2005PASJ...57S...1D}),
and the application to the near-infrared has extended these techniques 
to larger extinctions
\citep[e.g.][]{2001A&A...377.1023L}.
The extinction of individual 
background stars
are measured and converted to a column density of intervening gas and dust 
assuming appropriate properties for dust in molecular clouds. Studies of the
extinction law in nearby clouds show that the dust is best represented by models
with ratios of total to selective extinction, $R_V = 5.5$ 
(\citealt{2009ApJ...690..496C, 2012A&A...540A.139A, 2013A&A...549A.135A}).
For the nearby clouds imaged with Spitzer, 2\ee4 to 1\ee5 background stars were
identified \citep{2009ApJS..181..321E}; 
when added to the 2MASS data base, extinctions up to $A_V = 40$ mag can be measured
with spatial resolution of about 270\arcsec\
\citep{2010ApJ...723.1019H}. The conversion from $A_V$ to mass surface density
is 
\begin{equation}
\Sigma_{gas} ({\rm g\ cm^{-2}}) = \mu\,m_H \left[1.086\, C_{\rm ext}(V)\right]^{-1} A_V
\end{equation}
where $\mu = 1.37$. 
$C_{\rm ext}(V)$ is the extinction 
per column density of H nucleons,
$N(\mathrm{H}) = N(\mathrm{HI}) + 2 N(\mathrm{H}_2)$
\citep{2003ARA&A..41..241D}
(see on-line tables at
\url{http://www.astro.princeton.edu/~draine/dust/dustmix.html}).
Two different grain models are available for $R_V = 5.5$.
For normalized Case A grains, the newer models,
$C_{\rm ext}(V) =  6.715\ee{-22}$ cm$^{2}$, and
$\sigmagas = 15 \av$ \msunpc.
For Case B grains,
$C_{\rm ext}(V) =  4.88\ee{-22}$ cm$^{2}$, and
$\sigmagas = 21 \av$ \msunpc.
The Case B grains match observations well
\citep{2013A&A...549A.135A}, but have some
theoretical issues (Draine, personal communication).
Following 
\citet{2010ApJ...723.1019H}
we adopt the normalized Case A grain model, noting the possibility
that all masses and surface densities are about 40\% higher.
Note that $C_{ext} = 4.896\ee{-22}$ cm$^2$ (normalized Case A)
or $5.129\ee{-22}$ cm$^2$ (Case B)
for $R_V = 3.1$, so use of the diffuse ISM
conversions will result in higher 
estimates of \sigmagas. 
These values may apply to regions with $\av < 2$ mag, but
\citet{2013A&A...549A.135A} finds no clear change from
$R_V = 5.5$.

Ground-based submillimeter mapping of dust continuum emission provides
an alternative, but it requires assumptions about both the dust emissivity 
properties at long wavelengths and the dust temperature
distribution along the line of sight, usually taken to be constant. 
In addition, the ground-based submillimeter maps lose sensitivity to large-scale
structure.
The resolution, however, can be significantly higher than for extinction mapping, 
$\sim$ 10's of arcseconds rather than 100's, and the submillimeter emission 
remains optically thin even at extreme column densities.
For a description of planned maps with SCUBA-2, see
\citet{2005prpl.conf.8485J} and \citet{2007PASP..119..855W}

More recently Herschel has mapped a large fraction of the nearby molecular
clouds from the far infrared through the submillimeter. 
The resulting spectral energy distributions
add information about the mean dust temperature along the line of sight and
thus  yield more accurate column density maps of molecular clouds
(e.g., \citealt{2012A&A...540A..10s,2013MNRAS.432.1424K,2013ApJ...767..126S,2013ApJ...766L..17S}).
The spatial resolution of Herschel lies between the standard extinction map 
scale and the ground-based submillimeter maps.
We can expect a rich harvest of results once these surveys are fully
integrated. 

\bigskip
\noindent
\textbf{2.2
The Mass and Surface Density of Clouds}
\bigskip

The three largest clouds in the c2d project were completely mapped
at 1.1 mm using Bolocam on the CSO
(\citealt{2006ApJ...638..293E,2006ApJ...644..326Y,2007ApJ...666..982E}).
Additional maps were made at 850 \micron\ with SCUBA for
Ophiuchus
\citep{2004ApJ...611L..45J},
Perseus 
\citep{2008A&A...482..855H,2006ApJ...646.1009K},
and Orion 
\citep{1999ApJ...510L..49J,2006ApJ...653..383J}.
Smaller clouds and regions were mapped at 350 \micron\ by
\citet{2007AJ....133.1560W}
providing higher resolution data at those wavelengths than was possible
with Herschel.
A uniform reprocessing of {\it all} SCUBA data provides a data base for many individual
regions 
\citep{2006ASPC..356..275D}.

On larger scales, we now have
surveys of the Galactic Plane at mm/smm wavelengths,
notably the Bolocam Galactic Plane Survey (BGPS)
(\citealt{2011ApJS..192....4A,2010ApJS..188..123R,2013ApJS..208...14G})
and the APEX Telescope Large Area Survey of the Galaxy (ATLASGAL) projects
(\citealt{2009A&A...504..415S,2013A&A...549A..45C}).

The primary product of these blind submillimeter 
surveys was a catalog of relatively dense structures.
Although dust continuum emission is related to dust
temperature and column density throughout the cloud, the loss of sensitivity to large scale
emission for ground-based instruments results in an effective spatial filtering. 
This yields a particular sensitivity to small-scale column density enhancements and
thus picks out structures with high volume density.
For nearby clouds, mm sources correspond to cores, the sites of individual 
star formation 
\citep{2006ApJ...638..293E}, 
whereas Galactic Plane 
surveys mostly pick up clumps, the sites of cluster formation
\citep{2011ApJ...741..110D}.
This can also be seen in the fraction of cloud mass identified through the sub-mm mapping. In nearby clouds the sub-mm sources account for a few percent of the cloud mass \citep{2004ApJ...611L..45J,2006ApJ...646.1009K}, while at larger distances the fraction of cloud mass observed appears to be much higher, reaching tens of percent, and the slope of the clump mass function flattens 
\citep{2001ApJ...552..601K,2007ApJ...668..906M}.

For the nearby clouds, Spitzer data were used to distinguish protostellar
cores from starless cores, almost all of which are probably prestellar
(i.e. gravitationally bound),
using the definitions of
\citet{2007prpl.conf...17D} and \cite{2007prpl.conf...33W}.
This distinction allowed a clarification of the properties of prestellar cores.
Three results are particularly salient for this review.

First, the core mass function was consistent with a picture in which core masses
map into stellar masses with a core-to-star efficiency of 
$\eps = 0.25$ to 0.4
(\citealt{2008ApJ...684.1240E, 2010ApJ...710.1247S, 2007A&A...462L..17A,
2010A&A...518L.102A}).

Second, the timescale for prestellar cores with mean densities above
about $10^4$ \cmv\ to evolve into protostellar cores is about 0.5 Myr
\citep{2008ApJ...684.1240E}.
Third, prestellar cores are strongly concentrated
to regions of high extinction.
For example, 75\% of the cores lie above 
$A_V = 8$, 15, and 23 mag in Perseus, Serpens, and Ophiuchus, respectively
\citep{2004ApJ...611L..45J,2006ApJ...646.1009K,2007ApJ...666..982E}.
These were some of the
first quantitative estimates of the degree to which star formation
is concentrated to
a small area of the clouds characterized by high surface densities.
In contrast, less than 20\% of the area and 38\% of the cloud mass 
lies above $A_V = 8$ mag 
(Evans et al., submitted).

For 29 nearby clouds mapped by Spitzer, Evans et al. (submitted) have
measured mean surface densities above the $\av = 2$ contour. These have
a mean value of $\sigmagas = 79\pm 22$ \msunpc. For maps of 12 nearby
clouds that go down to $\av = 0.5$ mag with 0.1 pc resolution, 
mean $\av = 1.6\pm 0.7$ mag for star forming clouds and $\av = 1.5\pm0.7$ mag
if 5 non-star-forming clouds are included
\citep{2009A&A...508L..35K}.
Clearly, the mean surface densities are heavily influenced by 
how low an extinction contour is included.
These mean extinctions would correspond to 23 to 26 \msunpc\ with the
adopted extinction law, but other extinction laws
could raise these values by about 40\%.

\bigskip
\noindent
\textbf{2.3
Dust versus CO Observations}
\bigskip

The images of clouds from extinction or Herschel images of dust emission are
qualitatively different from earlier ones. At low extinction levels, clouds
appear wispy, windswept, and diffuse, more like cirrus clouds in our atmosphere
than like cumulus clouds. The Herschel images tracing higher extinction regions,
reveal a strong theme of filaments and strands with high contrast against the more
diffuse cloud (see chapter by Andr\'e et al.).
These modern images of cloud properties must be contrasted with older images,
which still dominate the mental images of molecular clouds for many astronomers.
The very definition of molecular cloud depends on the technique used to measure it.
For the nearby clouds, optical and near-infrared imaging define clouds to well 
below $A_V = 1$ and it is only at $A_V<1$ that log-normal column 
density distributions become apparent, as discussed in later sections, while
their SFR properties were generally measured down to only $A_V = 2$, 
as discussed in the next section.
In contrast, the maps of molecular clouds based on maps of CO probe down to a
particular value of antenna temperature or integrated intensity, and the conversion
to gas surface density depends on that limiting value. 

The most detailed study comparing extinction and CO at low levels is that by
\citet{2010ApJ...721..686P}, which combined deep extinction mapping with 0.14 pc
resolution with extensive CO maps
\citep{2008ApJ...680..428G}.
CO emission can be seen down to $\av \sim 0.1$ mag, but the conversion from
CO emission to mass surface density rises rapidly as \av\ decreases below
1 mag. 
The mass in areas where
\coo\ is not detected is roughly equal to the mass where it was detected.
The net result is a mean surface density of 39 \msunpc, based on the total
cloud mass and area \citep{2010ApJ...721..686P}.

For more distant clouds, or for clouds at low Galactic longitude and 
latitude, extinction and submillimeter maps cannot easily 
separate the cloud from the general Galactic field without 
additional kinematic data.  The main
tracer of clouds across the Galaxy has been maps of CO and isotopologues.
One of the earliest and most influential studies of cloud properties was by
\citet{1987ApJ...319..730S},
and this study remains the enduring image in the mental toolkit of many theorists
and nearly all extragalactic astronomers.
By defining clouds at a threshold of 3 K or higher in $T_R^*$ and assuming that
the virial theorem applied, 
\citet{1987ApJ...319..730S}
derived a mean surface density for the inner
Galaxy clouds of $\sigmagas = 170$ \msunpc, a value still quoted by many
astronomers. When scaled to the newer distance to the Galactic Center, this
value becomes 206 \msunpc\
\citep{2009ApJ...699.1092H}.
The Solomon et al.~work was however based on severely undersampled maps,
and the selection criteria favored ``warm" clouds with active star formation.
It has been superseded by better sampled and deeper maps of \coo\ by
\citet{2009ApJ...699.1092H}
and
\citet{2010ApJ...723..492R}.
\citet{2009ApJ...699.1092H} found that the median surface density of
inner Galaxy molecular clouds is 42 \msunpc\ within the extrapolated 1 K contour
of CO, but noted that abundance
variations could raise the value to 80-120 \msunpc.
\citet{2010ApJ...723..492R}
found a mean $\sigmagas = 144$ \msunpc\ within the $4\sigma$ contour of \coo,
which traced gas with a median $\av > 7$ mag, roughly. 
These values are thus measuring a fraction of the cloud with elevated surface 
densities, if those clouds are like the local clouds.
\citet{2010ApJ...723..492R} find that \sigmagas\ declines for Galactocentric
radii beyond 6.6 kpc to the lower values seen in nearby clouds.
These values depend on assumptions about CO abundance and excitation.
The definition and properties of clouds appear to depend on location in
the Galaxy as well as on technique and sensitivity.

Extensive maps in molecular lines of the nearby clouds have also 
allowed studies of the dynamical properties of the clouds  and locations of outflows
(e.g., \citealt{2006AJ....131.2921R}). These may also be used to compare to 
detailed simulations of line profiles from turbulent clouds.

\bigskip
\noindent \textbf{2.4
Measuring the SFR in MCs: Methods and Uncertainties}
\bigskip

Methods for measuring the SFRs in general have been
reviewed and tabulated by \citet{2012ARA&A..50..531K}.
The most direct measures of the SFRs in nearby 
molecular clouds are based on counting of young stellar objects (YSOs)
 and assigning
a timescale to the observed objects. Recent surveys of molecular
clouds with Spitzer provide a quite complete census of YSOs with 
infrared excesses \citep{2009ApJS..181..321E,2013AJ....145...94D}. 
There is however considerable uncertainty in the low luminosity range,
where contamination by background objects becomes severe. Tradeoffs
between completeness and reliability are inevitable. The YSO 
identifications in \citet{2009ApJS..181..321E} emphasized reliability;
other methods have suggested 30\% to 40\% more YSOs (e.g.,
\citealt{2012AJ....144...31K,2013ApJS..205....5H}).
Optical photometry and spectroscopy also provided valuable follow-up data
for the later stages (e.g.,
\citealt{2009ApJ...691..672O,2010A&A...513A..38S,2008ApJ...680.1295S}).
Although we may expect updates as the Herschel surveys of the Gould
Belt clouds are fully analyzed, initial results from the Orion
clouds suggest that the percentage of sources identifiable only with
Herschel data is about 5\% 
\citep{2013ApJ...767...36S}.

The selection of YSOs by infrared excess means that older objects
are not counted, but it has the advantage that the lifetime of infrared
excess has been studied extensively by counting the fraction of stars
with infrared excess in clusters of different ages. The half-life
of infrared excess is determined to be $\texcess = 2 \pm 1$ Myr
(e.g., \citealt{2009AIPC.1158....3M}),
with the main source of uncertainty being the choice of pre-main sequence 
(PMS) evolutionary tracks. 
All the measures of SFR (\sfr) discussed later depend
on the assumed half-life of infrared excesses. Increases in the ages of
young clusters (see chapter by Soderblom et al.) will decrease the estimates
proportionally.
The third element needed is the mean mass of a YSO, for which
we assume a fully sampled system IMF, with $\mean{\mstar} = 0.5$ \msun\
\citep{2003PASP..115..763C}.
The nearby clouds are not forming very massive stars, so they are not
sampling the full IMF; this effect would cause an overestimate of \sfr.

Putting these together,
\begin{equation}
\mean{\sfr} = {N(YSOs) \mean{\mstar} /t_{excess}}.
\label{ysocounteq}
\end{equation}
This equation has been used to compute \sfr\ for 20 nearby clouds 
\citep{2009ApJS..181..321E,2010ApJ...723.1019H}, with the assumed {\texcess} = 2 Myr being the major source of 
uncertainty \citep[for a numerical study investigating the effect of varying {\texcess}, see][]{2012ApJ...761..156F}.
From these measures of \sfr, one can compute surface densities of
star formation rate, \sigmasfr, efficiencies, defined by
$\sfe = M(YSO)/[M(YSO) + M(cloud)]$ \citep[see e.g., the definition in][]{2013ApJ...763...51F},
rates per mass (often referred to as efficiencies in extragalactic work),
defined by $\sfr/\mgas$, and their reciprocals, the depletion times, \tdep.
SFRs determined from star counting are the most
reliable, but are available only for nearby clouds
\citep[e.g.,][]{2010ApJ...723.1019H,2010ApJ...724..687L}
reaching out to Orion
\citep{2012AJ....144..192M}. However, even the Orion clouds are not fully representative of the regions of
massive star formation in the inner galaxy or in other galaxies.

Other methods of measuring \sfr\ are mostly taken from extragalactic
studies, where star counting is impractical, and their application to the Galaxy is tricky. Among the methods listed
in \citet{2012ARA&A..50..531K}, those that are not too sensitive to extinction, such as 24 \micron\ emission, 
total far-infrared emission, and thermal radio
continuum, may be useful also in localized regions of the Galactic Plane. 
\citet{2013ApJ...765..129V} tested a number of these and found that none
worked well for nearby regions where star counts are available. They noted
that all the extragalactic methods assume a well-sampled IMF and well-evolved cluster
models, so their failure in nearby regions is not surprising. They did find
consistency in estimated \sfr\ 
between thermal radio and total far-infrared methods for regions with
total far-infrared luminosity greater than $10^{4.5}$ \lsun.
As long as the overall SFR in a region
is dominated by massive, young clusters, the measures that are sensitive
to only the massive stars work to within a factor of two (e.g.,
\citealt{2011AJ....142..197C}).

In summary, methods using YSO counting are reasonably complete in nearby
($d < 500$ pc) clouds, and SFRs, averaged over 2 Myr,
are known to within a factor of about 2. For more distant clouds, where
YSO counting is not practical, indirect measures can fail by orders of 
magnitude unless the region in question has a sufficiently well sampled
IMF to satisfy the assumptions in models used to convert tracers into
masses of stars.

\bigskip
\noindent
\textbf{2.5
The Mass in Young Stars and the SFR in MCs}
\bigskip

The Spitzer/c2d survey summarized by \cite{2009ApJS..181..321E} mapped
seven nearby star-forming regions, Perseus, Ophiuchus, Lupus I, III, and
IV, Chamaeleon II, and Serpens. Additional 
data for JHK bands from the 2MASS project 
provided critical information for sorting
YSOs from background stars and galaxies in Spitzer surveys
(\citealt{2007ApJ...663.1149H} for c2d and Gould Belt,
\citealt{2010ApJS..186..259R} for Taurus, and
\citealt{2009ApJS..184...18G} for Orion and nearby clusters).

Using the near-infrared extinction maps,
these studies reveal a relatively small scatter in the star formation rate 
per unit area (\sigmasfr) of the c2d clouds. 
The values of \sigmasfr\ are in the range 0.65-3.2 M$_\odot$ Myr$^{-1}$~pc$^{-2}$, 
with an average of 1.6 M$_\odot$ Myr$^{-1}$~pc$^{-2}$. 
Efficiencies per 2 Myr, the characteristic
timescale for young stellar objects that can be detected and
characterized at these wavelengths, are generally small, 3\% to
6\%. This is similar to the star formation efficiency ($\sfe$) 
inferred from
2MASS studies of the Perseus, Orion A and B and MonR2 molecular clouds
\citep{2000AJ....120.3139C}. In the smaller group L673,
\cite{2010ApJ...725.2461T} likewise find 
$\sfe = 4.6$\%. 
In contrast, \cite{2003ARA&A..41...57L} concluded that the $\sfe$ 
is larger for embedded clusters than for MCs as a whole. 
Typical values for embedded
clusters are in the range 10--30\%, with lower values usually found for less evolved clusters.

These estimates largely differ because of the scales probed. For
example, considering the larger scale Ophiuchus and Perseus clouds,
\cite{2008ApJ...683..822J} found that the $\sfe$
was strongly dependent on the column density and regions considered:
the younger clusters such as L1688, NGC~1333 and IC~348 showed
values of 10--15\%, contrasting with the $\sfe$ of a few \% found
on larger scales \citep{2008ApJ...683..822J,2009ApJS..181..321E,2013ApJ...763...51F}.
Likewise, \cite{2011A&A...535A..77M} utilized the Herschel Space Observatory and
IRAM~30~m observations to infer $\sigmasfr\ = 23$ 
M$_\odot$~Myr$^{-1}$~pc$^{-2}$ for the Serpens South cluster
\citep[see
also][]{2008ApJ...673L.151G}. \citeauthor{2008ApJ...673L.151G} noted
that the YSO surface density of the Serpens South cluster was $>$430~pc$^{-2}$
within a circular region with a 0.2~pc radius, corresponding to a
higher $ \sigmasfr \approx$~700~M$_\odot$~Myr$^{-1}$~pc$^{-2}$ --
again likely reflecting the density of material there. Zooming-in on
the material just associated with dense cores, the gas+dust mass
becomes comparable to the mass of the young stellar objects -- an
indication that the efficiency of forming stars is high once material
is in sufficiently dense cores
\citep{2008ApJ...684.1240E,2008ApJ...683..822J}. 

In particular, \citet{2013ApJ...763...51F} measured the SFE in 29 clouds and cloud regions with a new technique based on the column-density power spectrum. They find that 
the $\sfe$ increases from effectively zero in large-scale HI clouds and
non-star-forming clouds to typical star-forming molecular clouds 
(\mbox{$\sfe=1\%$--$10\%$}) to dense cores ($\sfe>10\%$), 
where the SFE must eventually approach the core-to-star 
efficiency, \mbox{$\eps\approx0.3$--$0.7$}.
Undoubtedly, the SFE is not constant over time within a  cloud. 
The observed SFE must depend both on the properties of a cloud 
{\it and} on its the evolutionary state. The above qualitative SFE 
sequence encapsulates both of these effects.

\begin{figure*}[t]
\centerline{
\includegraphics[angle=-90,width=0.49\linewidth]{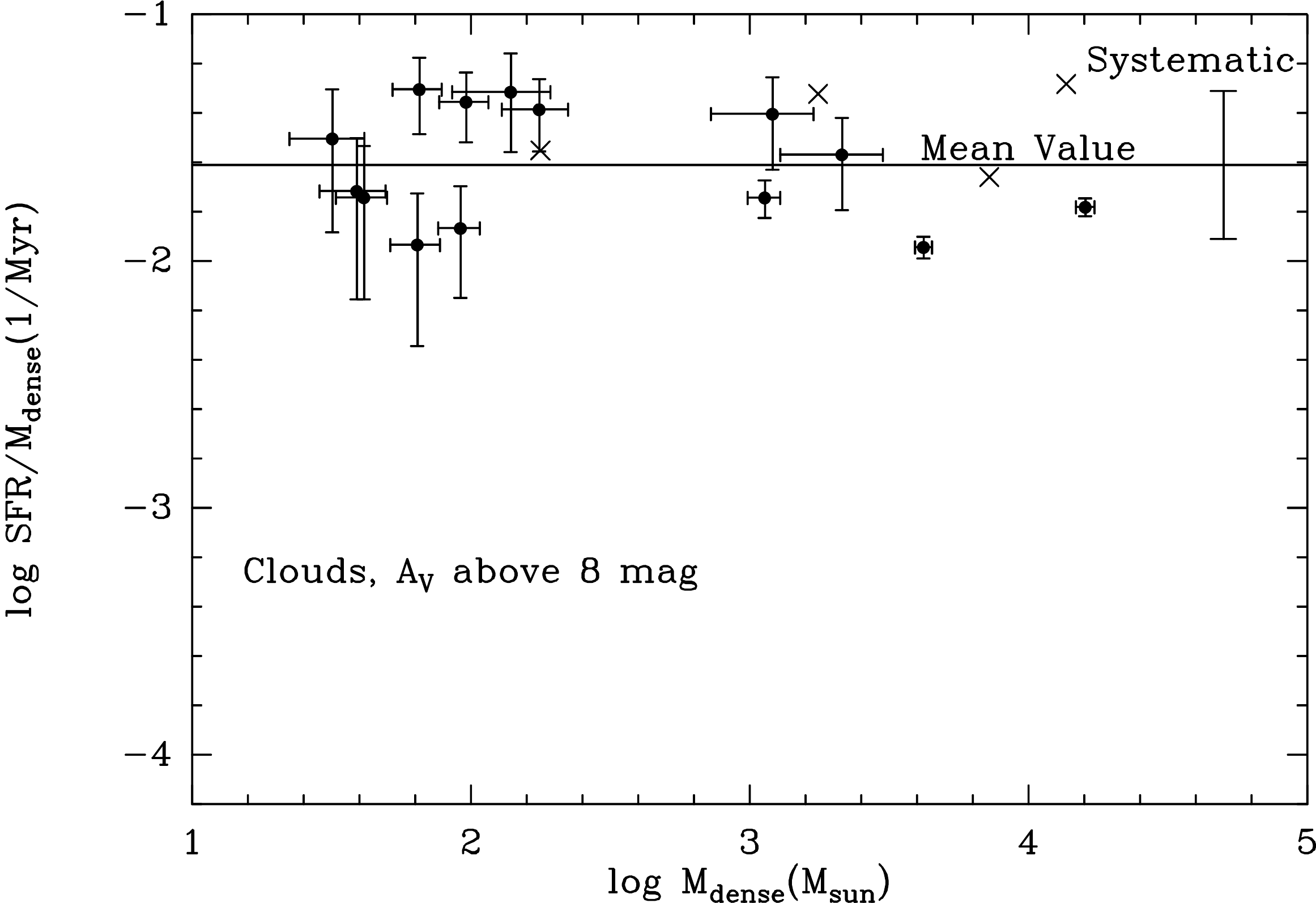}
\includegraphics[angle=-90,width=0.49\linewidth]{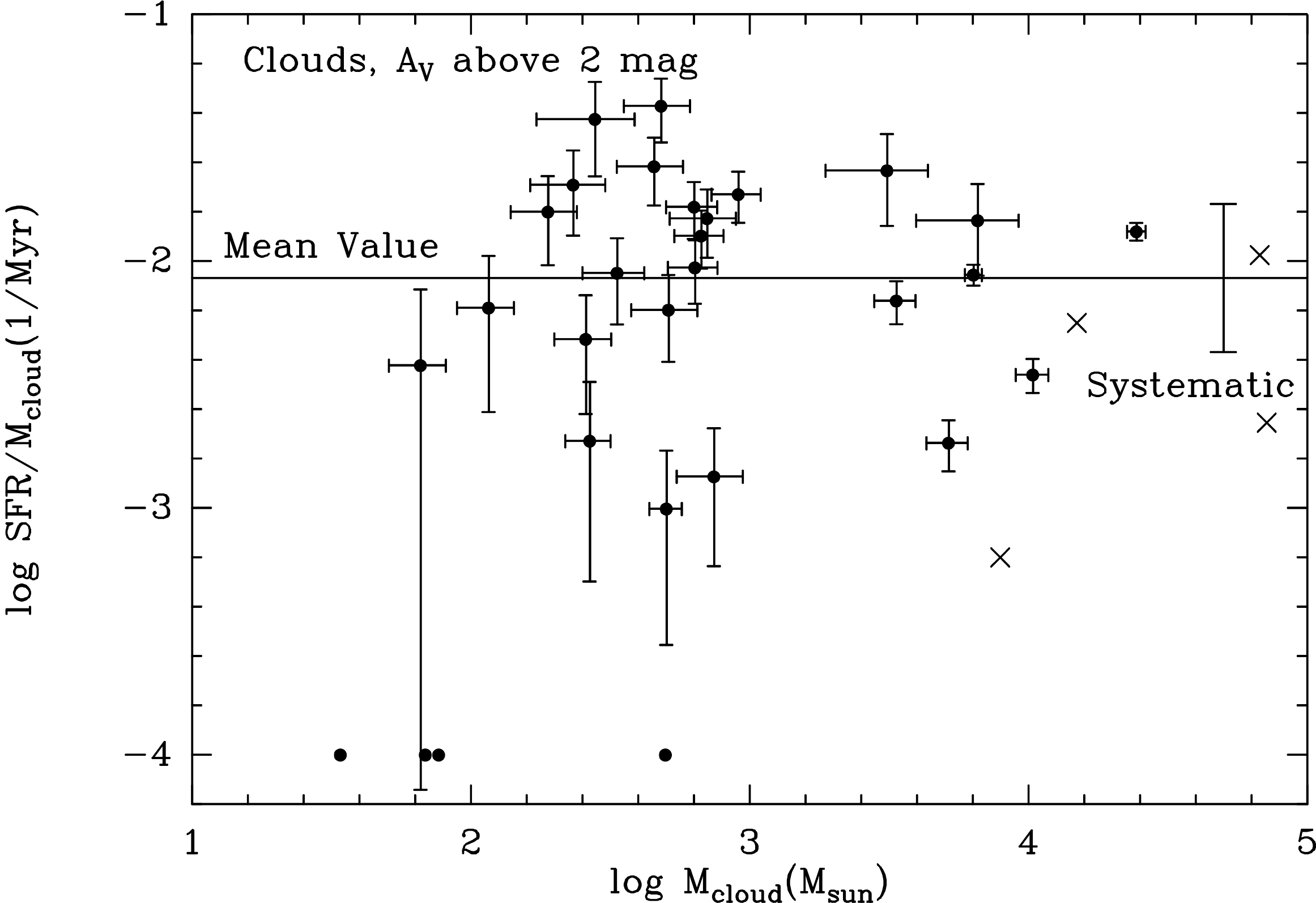}}
\caption{The log of the SFR per mass of dense gas versus
the log of the mass of dense gas (left panel) and 
the log of the SFR per total cloud mass versus the 
log of the total cloud mass (right panel). The filled circles are based on 
the c2d and Gould Belt clouds (Evans et al. submitted), while the crosses
represent Orion A, Orion B, Taurus, and the Pipe,
taken from 
\citet{2010ApJ...724..687L}. 
While there are some differences in identification and selection of
YSOs, they are small. On the left panel, the extinction contour defining the dense gas is
$\av = 8$ mag for clouds taken from Evans et al. and $A_K = 0.8$ mag
for those taken from 
\citet{2010ApJ...724..687L}. On the right panel, the extinction contour defining the cloud is usually
$\av = 2$ mag for clouds taken from Evans et al. and $A_K = 0.1$ mag
for those taken from
\citet{2010ApJ...724..687L}.
Uncertainties on observables, including cloud distance, have been propagated
for the c2d and Gould Belt clouds; the requisite information is not available
for the ``Lada" clouds.  The four points plotted at $-4$ on the y axis
are clouds with no observed star formation.
The horizontal lines show the mean values for the c2d and Gould Belt
clouds  and the error bars represent the likely {\it systematic} 
uncertainties, dominated by those in the SFR.
}
\label{fig:ladaratio}
\end{figure*}

In systematic comparisons between clouds or substructures of similar
scales, differences are also seen: \cite{2010ApJ...724..687L} combined
the data for a sample of MCs and found a strong (linear)
correlation between the SFR and the mass of the cloud
at extinctions above $A_K \approx 0.8$~mag ($\av \approx 7.5$ mag). 
\cite{2010ApJ...723.1019H}
and \cite{2011ApJ...739...84G} examined the relation between the SFR
and local gas surface density in subregions of clouds
and likewise found relations between the two:
\cite{2011ApJ...739...84G} found that 
$ \sigmasfr\ \propto \sigmagas^2$, while
\cite{2010ApJ...723.1019H} found 
a relation with a steeper increase (faster than
squared) at low surface densities and a less steep increase (linear) at
higher surface densities. One difference between these studies was 
the inclusion of
regions forming more massive stars: while the study by
\citeauthor{2011ApJ...739...84G} includes SFRs based
on direct counts for a number of clusters forming massive stars, the
\citeauthor{2010ApJ...723.1019H} study is based on the far-infrared luminosities and HCN column
densities for a sample of distant, more active, star-forming
regions, which may underestimate the true SFRs 
\citep{2010ApJ...723.1019H,2013ApJ...765..129V}.

Another result of these studies is the presence of a star formation
threshold in surface density: 
\cite{2010ApJ...723.1019H} and \cite{2010ApJ...724..687L}
find evidence for a characteristic extinction threshold
of $\av \approx 7$  to 8 mag
separating regions forming the great
majority of stars from those where star formation is rare. 
Similar thresholds have previously been suggested,
e.g., on the basis of mapping of C$^{18}$O emission
\citep{1998ApJ...502..296O} and (sub)millimeter continuum maps 
\citep[e.g.,][]{2004ApJ...611L..45J, 2006ApJ...646.1009K, 2007ApJ...666..982E} 
of a number of the nearby star-forming clouds. 
With our adopted conversion (see \S 2.1), the extinction threshold translates
to gas surface density of about 120$\pm$20~$M_\odot$~pc$^{-2}$, but
other conversions would translate to about 160 ~$M_\odot$~pc$^{-2}$ 
(see, e.g., \citealt{2013arXiv1309.7055L}).

All these methods measure the SFR 
for relatively nearby star-forming regions where individual young
stellar objects can be observed and characterized. An alternative on
larger scales is to measure the SFR using the
free-free emission, for example from the Wilkinson Microwave Anisotropy
Probe (WMAP). The free-free emission is expected to be powered by the
presence of massive stars through the creation of HII regions. Using
this method and by comparing to the GMC masses from CO line maps,
\cite{2011ApJ...729..133M} find varying star formation efficiencies
ranging from 0.2\% to 20\% with an average of 8\%. However, as the
author notes, the sample is biased toward the most luminous sources of
free-free emission, potentially selecting 
the most actively star-forming regions in our Galaxy. 

Observationally, the SFR scales with the cloud mass
and, with a much reduced scatter, on the mass of
``dense" gas 
(\citealt{2010ApJ...724..687L,2012ApJ...745..190L}).
Figure \ref{fig:ladaratio} shows the log of
the SFR/mass versus the mass, using either the mass of dense gas
($\av > 8$ mag), or the full cloud mass, down to $\av \sim 2$ mag.
The large scatter in the star formation rate per unit total gas mass 
and the much smaller scatter per unit mass of dense gas
are clearly illustrated.
The average values for the c2d plus Gould Belt clouds
are as follows (Evans et al. submitted):
$\tdep = 201\pm240$ Myr for clouds, $47\pm24$ Myr for dense gas;
$\tff = 1.47\pm0.58$ Myr for clouds, $0.71\pm 0.38$ Myr for dense gas;
$\sfrff \equiv \tff/\tdep = 0.018\pm 0.014$ for clouds, 
$0.018\pm0.008$ for dense gas. 

Removing this first-order dependence by looking at variations in SFR per
mass of either cloud or dense gas versus other, non-dimensional 
parameters is probably the best way forward. 
In doing so, we need to be clear about the entity in 
which we are comparing the SFRs. It could be for the
whole cloud, defined by either extinction or CO emission, most of which
lies well below the dense gas threshold, or it could refer only to 
the denser gas, for which the dispersion is much smaller. 
The latter might best be thought of as
the star forming portion of the cloud, which is often called a clump
\citep{2000prpl.conf...97W}.

\bigskip
\noindent
\textbf{2.6
Comparison with Extragalactic Results}
\bigskip

As many surveys of the Milky Way are becoming available, it is
increasingly interesting to put our Galaxy in the context of galactic
star formation studies. For other galaxies, the standard way to look
at star formation is via the Kennicutt-Schmidt (KS) relation
\citep{1998ApJ...498..541K,1998ARA&A..36..189K}.
Generalizing earlier suggestions \citep{1959ApJ...129..243S,1963ApJ...137..758S},
involving a relation between star formation and gas volume densities,
\citet{1998ApJ...498..541K} found a relation between more readily 
measured surface densities of the form:
\begin{equation}
\sigmasfr \propto \sigmagas^N
\end{equation}
with $N \approx 1.4$ providing a good fit over many orders of magnitude
for averages over whole galaxies.

Studies of resolved star formation over the face of nearby galaxies
reveal a more complex picture with a threshold around
$\sigmagas = 10$ \msunpc, roughly coincident with the transition from
atomic to molecular-gas dominated interstellar media, and a linear
dependence on the surface density of molecular gas, \sigmamol, above
that threshold \citep{2010AJ....140.1194B}.
Another transition around $\sigmamol = 100$ \msunpc, where the transition
from normal to starburst galaxies is found, has been suggested 
\citep{2010MNRAS.407.2091G,2010ApJ...714L.118D},
but this transition depends on interpretation of the CO observations, particularly the uncertainty in the
extrapolation of the total column density.
The existence of this second transition is consistent with relations between star
formation and dense gas measured by HCN emission found earlier by
\citet{2004ApJ...606..271G} and extended to clumps forming massive
stars in the Milky Way by \citet{2005ApJ...635L.173W}.
Together with the evidence for a sharp increase in SFR above
about $\sigmagas = 120$ \msunpc\ in local clouds,
these results led \citet{2012ARA&A..50..531K} to suggest
a second transition to a more rapid form of star formation
around $\sigmamol \approx 100 - 300$ \msunpc, depending on the
physical conditions.

When the nearby clouds are plotted on the KS relation, all but the
least active lie well above the relation for whole galaxies or the
linear relation found in nearby galaxies
\citep{2009ApJS..181..321E}. 
They are more consistent with
the relation for starburst galaxies and the dense gas relations noted above
\citep{2010ApJ...723.1019H}. \citet{2012ApJ...761..156F}, \citet{2012ApJ...745...69K}, and \citet{2013arXiv1307.1467F} provide suggestions to resolve this discrepancy between extra-galactic and Milky Way SFRs.

\section{\textbf{THEORY}}

The complexity of turbulence, especially in the supersonic regime and in the case of magnetized and self-gravitating gas,
precludes the development of fully analytical theories of the dynamics of MCs. However, the non-linear chaotic behavior
of turbulent flows results in a macroscopic order that can be described with universal statistics. These statistics are flow
properties that only depend on fundamental non-dimensional parameters expressing the relative importance of various
terms in the underlying equations, such as the Reynolds and Mach numbers.
Once established, scaling laws and probability density functions (PDFs) of turbulent flow variables can be applied to any
problem involving turbulent flows with known parameters. Furthermore, because of the chaotic nature of turbulent flows,
their steady-state statistics cannot even depend on the initial conditions. Because of this universality of turbulence, the 
statistical approach to the dynamics of MCs is a powerful tool to model the MC fragmentation induced by supersonic 
turbulence (with the obvious caveat that each MC is an individual realization of a turbulent flow, with specific local forces 
and possibly significant deviations from isotropy and steady state).

For many astrophysical problems involving supersonic turbulence, the turbulent fragmentation, that is the creation of
a highly nonlinear density field by shocks, is the most important effect of the turbulence. One can then tackle those
problems by focusing on the end result, the statistics of density fluctuations, largely ignoring the underlying dynamics.
The PDF of gas density in supersonic turbulence is then the statistics of choice. This is especially true in modeling star
formation, because the density enhancements, under characteristic MC conditions, are so large that the turbulence
alone can seed the gravitational collapse of compressed regions directly into the very non-linear regime. In other
words, much of the fragmentation process is directly controlled by the turbulence (the non-linear coupling of the
velocity field on different scales).

These considerations explain the choice of part of the numerical community, starting in the 1990's, to focus the
research on idealized numerical experiments of supersonic turbulence, using periodic boundary conditions,
isothermal equation of state, random driving forces, and, notably, no self-gravity 
\citep[e.g.][]{1998PhRvL..80.2754M,1999ApJ...526..279P,1998PhRvL..80.2754M,2004PhRvL..92s1102P,2007ApJ...665..416K,2010A&A...512A..81F}. 
These idealized simulations 
have stimulated important theoretical advances and better understanding of supersonic turbulence
\citep[e.g.][]{2010JFM...644..465F,2011PhRvL.106q4502A,2011PhRvL.107m4501G,2013arXiv1306.5768K,2013PhyD..247...54A}. 
The setup of these experiments (though with the inclusion of self-gravity and sink particles) are described in Section~4. 
Here we briefly summarize the most important numerical results regarding the gas density PDF (Section~3.1). We then introduce 
the concept of a critical density for star formation (Section~3.2), and discuss its application to the prediction of the SFR (Section~3.3).

\bigskip
\noindent
\textbf{3.1
The PDF of Gas Density in Supersonic Turbulence}
\bigskip

The gas density PDF was first found to be consistent with a log-normal function in three-dimensional simulations of 
compressible homogeneous shear flows by \citet{1993JFM...256..443B}, in two-dimensional simulations of supersonic 
turbulence by \citet{1994ApJ...423..681V}, and in three-dimensional simulations of supersonic turbulence by 
\citet{1997MNRAS.288..145P}. The log-normal function it is fully determined by its first and second order moments, the mean and the standard deviation, 
as it is defined by a trivial transformation of the Gaussian distribution: if the PDF of $\widetilde{\rho}=\rho/\rho_0$ 
is log-normal ($\rho_0$ is the mean density), then the PDF of $s={\rm ln}(\widetilde{\rho})$ is Gaussian:
\begin{equation}
p(s)ds=\frac{ 1 } { (2\pi\sigma_s^2)^{1/2}}\, {\rm exp} \left[ -\frac{(s - s_0)^2}{2\, \sigma_s^2} \right] ds \label{eq_pdf_rho} \vspace{0.1cm}
\end{equation}
where $s_0 = - \sigma_s^2 / 2$.
\citet{1997ApJ...474..730P} and \citet{1999intu.conf..218N} found that the standard deviation of the density scales linearly with the Mach number, $\sigma_{\widetilde{\rho}}= b \calm_{\rm s}$, where $\calm_{\rm s}\equiv \sigma_{\rm v}/c_{\rm s}$, and $b\approx 1/2$. This relation and the value of $b$ can be derived from a simple model based on a single shock \citep{2011ApJ...730...40P}. In terms of the logarithmic density, $s$, this is equivalent to,
\begin{equation}
\sigma_s^2 = {\rm ln}\left[1+b^2 \calm_{\rm s}^2\right] \label{eq_sig_rho}
\end{equation}

The log-normal nature of the PDF is valid only in the case of an isothermal equation of state (at low densities, time-averaging may be necessary to cancel out deviations due to large-scale expansions and retrieve the full log-normal shape). Strong deviations from the log-normal PDF can be found with a (non-isothermal) polytropic equation of state \citep{1998PhRvE..58.4501P,1998ApJ...504..835S,1999intu.conf..218N}. However, \citet{2010MNRAS.404....2G} and \citet{2012MNRAS.421.2531M} find nearly log-normal PDFs when a detailed chemical network including all relevant heating and cooling processes is taken into account. Further simulations tested the validity of eq.~(\ref{eq_sig_rho}) and derived more precise values of $b$ \citep[e.g.][]{1999ApJ...513..259O,2000ApJ...535..869K,2007ApJ...659.1317G,2007ApJ...665..416K,2008ApJ...688L..79F,2009A&A...494..127S,2010A&A...512A..81F,2011ApJ...727L..21P,2012ApJ...761..149K,2013arXiv1306.3989F}. Other numerical studies considered the effect of magnetic fields \citep[e.g.][]{2001ApJ...546..980O,2008ApJ...682L..97L,2011ApJ...730...40P,2012MNRAS.423.2680M}, and the effect of gravity \citep[e.g.][]{2000ApJ...535..869K,2008PhST..132a4025F,2011ApJ...731...59C,2011ApJ...727L..20K,2011MNRAS.410L...8C,2012ApJ...750...13C,2013ApJ...763...51F, 2013A&A...553L...8K}.

The value of $b$ depends on the ratio of compressional to total power in the driving force, with $b\approx 1$ for purely compressive driving, and $b\approx 1/3$ for purely solenoidal driving \citep{2008ApJ...688L..79F,2010A&A...512A..81F}. It has also been found that the PDF can be very roughly described by a log-normal also in the case of magnetized turbulence, with a simple modification of eq.~(\ref{eq_sig_rho}) derived by \citet{2011ApJ...730...40P} and \citet{2012MNRAS.423.2680M},
\begin{equation}
\sigma_s^2=\ln\left[1+b^2\mach^2 \beta/(\beta+1)\right]\,, \label{eq_sig_mhd}
\end{equation}
where $\beta$ is the ratio of gas to magnetic pressures, $\beta=8\pi\rho\cs^2/B^2 =2\cs^2/\va^2=2\macha^2/\mach^2$ for isothermal gas, $\va\equiv B/\sqrt{4\pi\rho}$ is the Alfv\'{e}n velocity, and $\calm_{\rm A}\equiv \sigma_{\rm v}/\va$ is the Alfv\'{e}nic Mach number. This relation simplifies to the non-magnetized case (eq.~(\ref{eq_sig_rho})) as $\beta \rightarrow \infty$. The derivation accounts for magnetic pressure, but not magnetic tension. In other words, it neglects the anisotropy of MHD turbulence that tends to align velocity and magnetic field. The anisotropy becomes important for a large magnetic field strength, and the relation (\ref{eq_sig_mhd}) is known to break down for trans- or sub-Alfv\'{e}nic turbulence, requiring a more sophisticated model.

It has been found that self-gravity does not affect the density PDF at low and intermediate densities, and only causes the appearance of a power-law tail at large densities, reflecting the density profile of collapsing regions \citep{2000ApJ...535..869K,2005ApJ...630..238D}. \citet{2011ApJ...727L..20K} have used a very deep AMR simulation covering a range of scales from 5 pc to 2 AU \citep{2005ApJ...622L..61P}, where individual collapsing cores are well resolved, and the density PDF develops a power law covering approximately 10 orders of magnitude in probability. By comparing the density profiles of the collapsing cores with the power-law high-density tail of the density PDF, they demonstrated that the PDF tail is consistent with being the result of an ensemble of collapsing regions. They point out that a spherically symmetric configuration with a power-law density profile, $\rho=\rho_0(r/r_0)^{-n}$ has a power-law density PDF with slope $m$ given by $m=-3/n$, and a projected density PDF with slope $p$ given by $p=-2/(n-1)$ (see also \citet{2010MNRAS.408.1089T}
and \citet{2013arXiv1310.4346G}). They find that in their simulation the power-law exponent decreases over time, and may be nearly stationary at the end of the simulation, when it reaches a value $m=-1.67$ for the density PDF, and $p=-2.5$ for the projected density PDF. Both values give $n=-1.8$, consistent with similarity solutions of the collapse of isothermal spheres \citep{1985MNRAS.214....1W}. \citet{2011ApJ...731...59C} find a very similar slope, $m=-1.64$, also in the case of magnetized turbulence, and \citet{2013ApJ...763...51F} show that this result does not depend on the Mach number or on the ratio of compressive to total power, consistent with being primarily the effect of self-gravity in collapsing regions. The slope has a slight dependence on the turbulence Alfv\'{e}nic Mach number, as the absolute value of $m$ increases with increasing magnetic field strength \citep{2012ApJ...750...13C}.

\bigskip
\noindent
\textbf{3.2
The Critical Density for Star Formation}
\bigskip

Current theories assume that molecular clouds are supersonically turbulent and that pre-stellar cores arise as gravitationally
unstable density fluctuations in the turbulent flow. As explained in Section~3.3, they derive the SFR as the mass fraction
above an effective critical density for star formation, $\rho_{\rm crit}$, which can be computed by assuming that the density PDF
is log-normal, as discussed in Section~3.1. The derivation of such a critical density is therefore a crucial step in current SFR models. 
\citet{2012ApJ...761..156F} have summarized the various treatments of the critical density in recent
theories of the SFR by \citet[][KM]{2005ApJ...630..250K}, \citet[][PN]{2011ApJ...730...40P}, and 
\citet[][HC]{2011ApJ...743L..29H,2013ApJ...770..150H}. Table~\ref{tab:theories} summarizes the main differences in those theories, in particular 
the different choice of density threshold. 

One of the fundamental dimensionless parameters in the theories is the virial parameter, given by the ratio of turbulent and gravitational energies.
Because of the difficulty of estimating the virial parameter in non-idealized systems, we distinguish between the theoretical virial parameter,
$\avirt=2E_\mathrm{kin}/E_\mathrm{grav}$, and the observational one:
\beq
\aviro=5\sod^2 R/GM,
\label{alpha_obs}
\eeq
where $E_\mathrm{kin}$ and $\sod$ include thermal effects.
The two virial parameters are equal for an isothermal, spherical cloud of constant density \citep{1992ApJ...395..140B, 2012ApJ...761..156F}; 
for centrally concentrated clouds, $\aviro>\avirt$. \citet{2012ApJ...761..156F} find that the fractal structure \citep[e.g.,][]{1990ASSL..162..151S,2009ApJ...692..364F} of the clouds and the assumed boundary conditions (particularly periodic vs.~non-periodic) can 
lead to order-of-magnitude differences between $\aviro$ and $\avirt$.
In the absence of a magnetic field, isolated clouds with $\avirt<1$ cannot be in equilibrium and
will collapse. Including the effects of surface pressure, the critical value of $\avirt$ for collapse is somewhat
greater than 1 \citep{1992ApJ...395..140B}.
The corresponding critical value for $\aviro$ depends on the density distribution;
for example, for a thermally supported Bonnor-Ebert sphere, it is 2.054. In the following, we refer to the virial parameter on the cloud diameter scale $L=2R$ as $\alpha_\mathrm{cl}$.

KM assumed that low-mass pre-stellar cores are thermally supported, so that in a turbulent
medium the Jeans length of a critical core is proportional to the 1D sonic length,
$\lambda_{\rm J,\, crit}=\ell_{\rm s,1D}/\phi_x$, where $\phi_x$ is a numerical factor that was
assumed to be of order unity. They derived the sonic length using a linewidth-size relation of the form $\sigma_{\rm v,1D}\propto r^q$. 
In their numerical evaluation, they adopted $q\simeq 1/2$, consistent with both simulations
\citep[e.g.][]{1995MNRAS.277..377P, 2007ApJ...665..416K, 2010A&A...512A..81F, 2013arXiv1306.3989F} and observations
\citep[e.g.][]{2002A&A...390..307O,2004ApJ...615L..45H, 2006ApJ...653L.125P,2009ApJ...707L.153P,2011ApJ...740..120R}. 
As a result, they obtained
\beq
\rho_{\rm crit,KM}=\left(\pi^2\phi_x^2/45\right)\alpha_{\rm cl}\calm_{\rm s,cl}^2 \, \rho_\cl ,
\label{rho_crit_KM}
\eeq
where $\alpha_{\rm cl}$ is the observational virial parameter for the cloud,
$\rho_\cl$ is the mean density of the cloud, and $\calm_{\rm s,cl}$, like all Mach numbers in this paper,
is the 3D Mach number.
They estimated the factor $\phi_x$ to be 1.12 from comparison with
the simulations of \citet{2003ApJ...585L.131V}. If the sonic length is defined based on the 3D velocity dispersion, then the coefficient in Equation~(\ref{rho_crit_KM}) changes from $(\pi^2\phi_x^2/45)$ to $(\pi^2\phi_x^2/5)$ \citep{2012ApJ...761..156F}. The KM result for $\rho_\mathrm{crit}$ is
a generalization of the result of \citet{1995MNRAS.277..377P} that $\rho_{\rm core,\,crit}\simeq \calm_{\rm s,\,cl}^2\rho_\cl$.

In PN, the critical density was obtained by requiring that the diameter of a critical Bonnor-Ebert sphere be equal to the characteristic thickness 
of shocked layers (identified as the size of prestellar cores), inferred from isothermal shock
conditions (MHD conditions when the effect of the magnetic field is included):
\beq
\rho_{\rm crit,PN} \simeq 0.067\, \theta_{\rm int}^{-2} \alpha_{\rm cl} {\cal M}_{\rm s,cl}^2\, \rho_\cl ,
\label{rho_crit_PN}
\eeq
where $\theta_{\rm int}=$
integral-scale radius/cloud diameter $\approx 0.35$. 
PN did not assume a velocity-size relation, so their result does not depend on a specific value of $q$. However, for a reasonable value of 
$q=1/2$, the critical density in KM agrees with that in PN to within a factor 2.

The virial parameter can also be expressed as $\alpha_{\rm cl} \propto \sigma_{\rm v,cl}^2/\rho_{\rm cl} R_{\rm cl}^2$. Thus, if the exponent 
of the line width-size relation is $q=1/2$, the critical density from equations (\ref{rho_crit_KM}) and
(\ref{rho_crit_PN}) becomes independent of cloud density, virial parameter and Mach number:
\begin{equation}
n_{\rm H,\, crit}\approx (5-10)\times 10^4 \,\, {\rm cm}^{-3} ( T_{\rm cl} / 10\,{\rm K} )^{-1},
\label{rho_crit_const}
\end{equation}
assuming $\sod=  0.72$~km\,s$^{-1}$\,pc$^{-1/2} (R/1\,$pc$)^{1/2}$ \citep{2007ARA&A..45..565M}.
This previously overlooked result suggests a possible theoretical explanation for the nearly constant
threshold density for star formation suggested by observations (see Sections~2 and 5.3).
However, it must be kept in mind that $n_{\rm H,\, crit}$ varies as the square of the coefficient in the
linewidth-size relation, so regions with different linewidth-size relations should have different critical densities.
It should also be noted that magnetic fields tend to reduce the critical density by an uncertain amount (see below),
and that the HC formulation of the critical density does not yield a constant value.

If the core is turbulent, 
{\it McKee et al.} (2014, in preparation) show that the critical density can be expressed as
\beq
\rho_{\rm core,\,crit}\simeq \left(\frac{\sigma_{\rm v,cl}}{\sigma_{\rm v,core}}\right)^2 \frac{\avircl}{\alpha_{\rm crit}}\; \rho_\cl .
\label{rho_crit}
\eeq
The numerical factor
\def\calo{{\cal{O}}}
$\alpha_{\rm crit}=\calo(1)$ must be determined from theory or simulation, but
the virial parameter of the cloud, $\alpha_{\rm cl}$, can be determined from observations.
Cores that are turbulent can be subject to further fragmentation, but this is suppressed by protostellar
feedback (e.g., \citealp{2013ApJ...766...97M}). 
For cores that are primarily thermally supported ($\sigma_{\rm v,1D,core}\simeq \cs$), this expression
has the same form as the KM and PN results above; if the core is turbulent, the critical density is reduced.

HC include the effects of turbulence in pre-stellar cores, but they
argue that their model 
does not have a critical density for star formation. That is true of their derivation of the stellar IMF, where cores of
different densities are allowed to collapse. For example, large
turbulent cores with density lower than that of thermally supported cores
are at the origin of massive stars, as in
\citet{2003ApJ...585..850M}. However, when integrating the mass spectrum of gravitationally unstable density fluctuations to derive the
SFR, HC face the problem of choosing the largest mass cutoff in the integral, 
corresponding to the most massive star,
which introduces an effective critical density.
They chose to extend the integral up to a fraction, $y_{\rm cut}$, of the cloud size, $L_{\rm cl}$. Their critical density is thus the density
such that the Jeans length (including turbulent support) is equal to $y_{\rm cut} L_{\rm cl}$:
\beq
\rho_{\rm crit,HC}=\left[\left(\frac{\pi^2}{5}\right) \,y_{\rm cut}^{-2} \, \calm_{\rm s,cl}^{-2} \,+ \left(\frac{\pi^2}{15}\right)\, y_{\rm cut}^{-1}\right]\, \alpha_{\rm cl}\,\rho_\cl\ ,
\label{rho_crit_HC}
\eeq
where the first term is the contribution from thermal support, and the second term the contribution from turbulent support.
The second term is larger than the first one if $\calm_{\rm s,cl}> (3/y_{\rm cut})^{1/2}$\,, and
can be obtained from equation (\ref{rho_crit}) by noting that, for $q=\frac 12$, 
one has $\calm_{\rm s,\, core,\, cut}^2/\calm_{\rm s, cl}^2=R_{\rm core,\, cut}/R_{\rm cl}
=y_{\rm cut}$ for the maximum core mass; the second term of equation (\ref{rho_crit_HC}) then follows
by taking $\alpha_{\rm core}=15/\pi^2$, similar to KM. HC assume that $y_{\rm cut}$
is a constant, independent of the properties of the cloud; the physical justification for
this has yet to be provided.
In summary, the critical density for KM and PN is defined for thermally supported cores, whereas that for HC is for the turbulent core 
corresponding to the most massive star.

Besides affecting the density PDF in a complex way (see Section 3.1), magnetic fields can alter the critical density. Equation (\ref{rho_crit}) remains valid, but
the values of the virial parameters change.
\citet{2003ApJ...585..850M} estimated that $\aviro$ is reduced by a factor $(1.3+1.5/\macha^2)$ in the presence of a magnetic field, provided that the gas is highly supersonic. If the core is embedded in a cloud with a similar value of $\macha$ and the cloud is bound, then the magnetic field
reduces $\aviro$ for both the core and the cloud, and the net effect of the field is small. \citet{2012MNRAS.423.2016H} finds
that the critical density depends on the ratio of $(\cs^2+\sigma^2+\va^2)$ on the cloud and core scales, so the field does
not have a significant effect if $\macha$ is about the same on the two scales.
By contrast, \citet{2012ApJ...761..156F} include magnetic fields in the KM derivation of $\rho_\mathrm{crit}$ by generalizing the comparison of Jeans length and sonic length to the comparison of the magnetic Jeans length with the magneto-sonic length. This reduces the KM value by a factor 
$(1+\beta^{-1})$, where
$\beta$ is the ratio of thermal to magnetic pressure, in agreement with the PN model (see Table~\ref{tab:theories}). 
For the HC model, instead, \citet{2012ApJ...761..156F} find that the critical density increases with increasing magnetic field strength,
because it depends inversely on $\calm_{\rm s, \, cl}$ (see eq.~\ref{rho_crit_HC}) rather than directly, as in the KM and PN models. These contradictory results on the effect of the magnetic field are discussed in more detail in
{\it McKee et al.} (2014, in preparation).

\begin{table*}
{\scriptsize
\caption{Six Analytical Models for the Star Formation Rate per Freefall Time.}
\label{tab:theories}
\def\arraystretch{1.5}
\setlength{\tabcolsep}{0pt}
\begin{tabular}{ p{0.11\linewidth} p{0.13\linewidth} p{0.08\linewidth} p{0.015\linewidth} p{0.24\linewidth} p{0.42\linewidth}}
\hline
\hline
Analytic Model & Freefall-time \mbox{Factor}  & \multicolumn{3}{p{0.34\linewidth}}{\mbox{Critical Density} \mbox{$\rhocrit/\meanrho=\exp(\scrit)$}} & $\sfrff$\\
\hline
KM                      & 1                                                & $(\pi^2/45)\,\phix^2$     &$\times$&  $\avircl\mach^2\left(1+\beta^{-1}\right)^{-1}$ & $\eps/(2\phit) \left\{1+\mathrm{erf}\left[(\sigs^2-2\scrit)/(8\sigs^2)^{1/2}\right]\right\}$ \\
PN                      & $\tff(\meanrho)/\tff(\rhocrit)$  & $(0.067)\,\theta^{-2}$ &$\times$&  $\avircl\mach^2 f(\beta)$ & $\eps/(2\phit) \left\{1+\mathrm{erf}\left[(\sigs^2-2\scrit)/(8\sigs^2)^{1/2}\right]\right\} \exp\left[(1/2)\scrit\right]$ \\
HC                      & $\tff(\meanrho)/\tff(\rho)$       & $(\pi^2/5)\,\ycut^{-2}$ &$\times$&  $\avircl\mach^{-2}\left(1+\beta^{-1}\right) + \tilde{\rho}_\mathrm{crit,turb}$ & $\eps/(2\phit) \left\{1+\mathrm{erf}\left[(\sigs^2-\scrit)/(2\sigs^2)^{1/2}\right]\right\} \exp\left[(3/8)\sigs^2\right]$ \\
multi-ff KM         & $\tff(\meanrho)/\tff(\rho)$       & $(\pi^2/5)\,\phix^2$     &$\times$&  $\avircl\mach^2\left(1+\beta^{-1}\right)^{-1}$ & $\eps/(2\phit) \left\{1+\mathrm{erf}\left[(\sigs^2-\scrit)/(2\sigs^2)^{1/2}\right]\right\} \exp\left[(3/8)\sigs^2\right]$ \\
multi-ff PN         & $\tff(\meanrho)/\tff(\rho)$       & $(0.067)\,\theta^{-2}$  &$\times$&  $\avircl\mach^2 f(\beta)$ & $\eps/(2\phit) \left\{1+\mathrm{erf}\left[(\sigs^2-\scrit)/(2\sigs^2)^{1/2}\right]\right\} \exp\left[(3/8)\sigs^2\right]$ \\
multi-ff HC         & $\tff(\meanrho)/\tff(\rho)$       & $(\pi^2/5)\,\ycut^{-2}$ &$\times$&  $\avircl\mach^{-2}\left(1+\beta^{-1}\right)$ & $\eps/(2\phit) \left\{1+\mathrm{erf}\left[(\sigs^2-\scrit)/(2\sigs^2)^{1/2}\right]\right\} \exp\left[(3/8)\sigs^2\right]$ \\
\hline
\end{tabular}
\\
}
\textbf{Notes.} 
Note that the critical density for the KM and multi-ff KM models were derived in \citet{2005ApJ...630..250K} based on the 1D velocity dispersion, while \citet{2012ApJ...761..156F} used the 3D velocity dispersion, which changes the coefficient from $(\pi^2/45)\phi_x^2$ to $(\pi^2/5)\phi_x^2$.
The function $f(\beta)$, entering the critical density in the PN and multi-ff PN models is given in \citet{2011ApJ...730...40P}. The added turbulent contribution $\tilde{\rho}_\mathrm{crit,turb}$ in the critical density of the HC model is given in \citet{2011ApJ...743L..29H}, \citet{2012ApJ...761..156F}, and Eq.~(\ref{rho_crit_HC}). For further details, see \citet{2012ApJ...761..156F}, from which this table was taken.
\end{table*}

\bigskip
\noindent
\textbf{3.3
The Star Formation Rate}
\bigskip

The dimensionless
{\it star formation rate per free-fall time}, SFR$_{\rm ff}$, is defined as
the fraction of cloud (or clump) mass converted into stars
per cloud {\it mean} free-fall time, $ t_{\rm ff,0} = \sqrt{3\,\pi/32\,G \rho_0}$, where $\rho_0$ is
the mean density of the cloud:
\begin{eqnarray}
{\rm SFR_{ff}}(t) = {\dot{M}_*(t)\over M_{\rm cl}(t)} \, t_{\rm ff,0},
\label{sfr_def}
\end{eqnarray}
where all quantities are in principle time-dependent.

Detailed comparisons between the KM, PN and HC models have been presented in \citet{2011ApJ...743L..29H} and in
\citet{2012ApJ...761..156F} and will only be summarized here. A more general formalism
to derive the SFR has been developed by \citet{2013MNRAS.430.1653H}, but as this author does
not provide explicit formulae for the SFR, it is difficult to include here the results of his model.

The common point between these models is that they all rely on the so-called turbulent fragmentation scenario for star formation,
with star formation resulting from a field of initial density fluctuations imprinted by supersonic turbulence. 
The models assume that gravitationally unstable density fluctuations are created by the supersonic turbulence and
thus follow the turbulence density PDF (eq.~(\ref{eq_pdf_rho})).
The SFR is then derived as the integral of the density PDF above a critical density (the HC model is actually derived from the integral
of the mass spectrum of gravitationally unstable density fluctuations, but, after some algebra and simplifying assumptions, it is reduced to an integral
of the density PDF above a critical density). The differences between the SFR models result from the choice of the critical density (see Section~3.2 and Table~\ref{tab:theories}),
at least when the same general formulation for the SFR is used for all models, as first proposed by \citet{2011ApJ...743L..29H} and
further explored by \citet{2012ApJ...761..156F}:
\begin{equation}
{\rm SFR_{ff}} = \frac{\eps}{\phit}\mathlarger{\int}_{\widetilde{\rho}_{\rm crit}}^{\infty}{\frac{t_{\rm ff,0}}{t_{\rm ff}(\widetilde{\rho})} \, p (\widetilde{\rho}) \,  \deriv \widetilde{\rho}}\,,
\label{sfr_int}
\end{equation}
where $p(\widetilde{\rho}) \deriv \widetilde{\rho} = \widetilde{\rho} p(s) ds$, and $p(s)ds$ is given by eq.~(\ref{eq_pdf_rho}),
with the variance following the relations (\ref{eq_sig_rho}) or (\ref{eq_sig_mhd}). 
Notice that, in the HC model, the relations (\ref{eq_sig_rho}) or (\ref{eq_sig_mhd})
include a scale dependence reflecting the property of the turbulence velocity power spectrum, $\sigma_s(R)=\sigma_{s,0}\,f(R)$
(e.g. eqn. (5) in \citet{2008ApJ...684..395H}). As mentioned below, this point is of importance when calculating the SFR.

In (\ref{sfr_int}), the PDF is divided by the free-fall time of each density, because
unstable fluctuations are assumed to turn into stars in a free-fall time. In steady state, the shape of the density PDF should be constant with time and thus
the turbulent flow must replenish density fluctuations of any amplitude. This may require a time longer that the free-fall time at a given density, hence
the replenishment factor, $\phit$, is introduced. It may in general be a function of $\widetilde{\rho}$, but given our poor understanding of
the replenishment process, it is simply assumed to be constant.  \citet{2011ApJ...743L..29H} suggested to use the turbulence turnover time as an estimate
of the replenishment time. They showed (in their Appendix) that this choice yields $\phit \approx 3$. KM assumed that the
replenishment time was of order the global free-fall time, not the local one, and their factor $\phit$ was defined accordingly. PN assumed
that the replenishment time was of order the free-fall time of the critical density, and did not introduce a parameter $\phit$ to parameterize the uncertainty of
this choice. Fitting each theory to numerical simulations, \citet{2012ApJ...761..156F} find that \mbox{$\phit\approx2$--$5$} in the best-fit models (see Figure~\ref{fig:model_comparison} below).

Note that the factor $t_{\rm ff,0}/\tff(\rho)$ appears \emph{inside} the integral~(\ref{sfr_int}) because gas with different densities has different freefall times,
which must be taken into account in the most general case \citep[see][]{2011ApJ...743L..29H}. Previous estimates for $\sfrff$ either used a factor $t_{\rm ff,0}/t_{\rm ff,0}=1$ \citep{2005ApJ...630..250K}, or a factor $t_{\rm ff,0}/\tff(\rhocrit)$ \citep{2011ApJ...730...40P}, both of which are independent of density and were thus pulled out of the general integral, Equation~(\ref{sfr_int}). We refer to the latter models as `single freefall' models, while Equation~(\ref{sfr_int}) is a `multi-freefall' model, but we can still distinguish three different density thresholds $\scrit=\ln(\widetilde{\rho}_{\rm crit})$ based on the different assumptions in the KM, PN, and HC theories. The general solution of Equation~(\ref{sfr_int}) is
\begin{equation} \label{eq:sfrff_basicsolution}
\sfrff = \frac{\eps}{2\phit} \exp\left(\frac{3}{8}\sigs^2\right) \left[1+\mathrm{erf}\left(\frac{\sigs^2-\scrit}{\sqrt{2\sigs^2}}\right)\right],
\end{equation}
and depends only on $\scrit$ and $\sigs$, which in turn depend on the four basic, dimensionless parameters $\alphavir$, $\mach$, $b$, and $\beta$, as shown and derived for all models in \citet{2012ApJ...761..156F} (see Table~\ref{tab:theories} for a summary of the differences between each theoretical model for the SFR).
It must be stressed that integrating analytically eq. (\ref{sfr_int}) to yield eq. (\ref{eq:sfrff_basicsolution}) is
possible {\it only} if the variance of the PDF does not entail a scale dependence. In the HC formalism, this corresponds to what these authors refer to 
as their ``simplified" multi-freefall theory (section 2.4 of \citet{2011ApJ...743L..29H}), while their complete theory properly accounts for the scale dependence.

The coefficient $\epsilon$ accounts for two efficiency factors: the fraction of
the mass with density larger than $\rho_{\rm crit}$ that can actually collapse (a piece of very dense gas may be too small or too turbulent to collapse)
and the fraction that ends up into actual stars (the so called core-to-star efficiency). The first factor should already be accounted for in the HC derivation of
eq.~({\ref{sfr_int}}), and therefore should not be incorporated into $\epsilon$. However, given the simplifications in the derivation, and the uncertainty in
choosing the extreme of integration of the mass spectrum mentioned above (\S~3.2), it may be necessary to allow for a fraction of this efficiency factor to be
included in $\epsilon$. The second factor (the core-to-star efficiency) is partly constrained by observations as well as analytical calculations and simulations,
suggesting a value $\epsilon \simeq 0.3-0.7$ \citep[e.g.][]{2000ApJ...545..364M,2012ApJ...761..156F,2013ApJ...763...51F}.
As in the case of $\phit$, $\eps$ is assumed to be a constant, even if, in principle, it may have a density dependence.
To be consistent with their physical meaning, the values of these two parameters should satisfy the conditions $\phit \ge 1$ and $\eps \le 1$.

As first proposed by \citet{2011ApJ...743L..29H}, the KM and PN models should be extended based on eq.~(\ref{sfr_int}), which is easily
done by inserting the KM and PN expressions for $\rho_{\rm crit}$ in that equation. The original KM and PN models were equivalent to
a simplified version of eq.~(\ref{sfr_int}), where one substitutes $t_{\rm ff}(\widetilde{\rho})$ with $t_{\rm ff,0}$ (KM) or $t_{\rm ff}(\widetilde{\rho}_{\rm crit})$
(PN). The effect of this extension is significant in the case of the KM model, because it introduces a dependence of SFR$_{\rm ff}$ on $\calm_{\rm s}$
that is otherwise missing, but rather small in the case of the PN model, where that dependence was already present.
\citet{2012ApJ...761..156F} showed that this extension from a single-freefall to a multi-freefall theory of the SFR does improve the match of the KM and PN models with their simulations (see \S~4.4).

\section{\textbf{SIMULATIONS}}

Numerical simulations where physical conditions may be easily varied and controlled, can greatly 
contribute, both qualitatively and quantitatively, to our understanding
of the most important factors influencing the star formation rate.
In this Section, we first discuss numerical methods, then present
results from large parameter studies.

\bigskip
\noindent
\textbf{4.1
Numerical Methods}
\bigskip

SFRs measured in numerical experiments should be primarily
influenced by the fundamental numerical parameters, such as
Mach numbers, virial numbers, types of external driving, etc.  However,
shortcomings in the numerical representation of turbulence, on the one hand,
and specific details (or lack thereof) in the recipes used to implement
accretion onto `sink particles' representing the real stars may also influence
the results significantly.

Some of the general shortcomings, particularly in relation to the need to
reproduce a turbulent cascade covering a wide range of scales, were reviewed in
\cite{2007prpl.conf...99K}. 
Since then, comparisons of numerical codes have been presented by
\cite{2009A&A...508..541K} 
for the case of decaying supersonic HD turbulence, and by
\cite{2011ApJ...737...13K} 
for the case of decaying supersonic MHD turbulence.
\cite{2010MNRAS.406.1659P} 
compared grid and particle methods for the case of driven, supersonic HD
turbulence.

From the point of view of the current discussion, the bottom line of these
comparisons is that the effective resolution of the various grid-based codes
used in the most recent SFR measurements
are quite similar, and that a much larger range of resolution results from the
use (or not) of AMR than from the use of different codes.  On problems where
direct comparisons can be made (mainly HD problems) grid- and particle-based
codes also show similar results. Even though
these problems do not have analytical solutions that may be used for
verification, the general behavior of both HD and MHD turbulence is well
represented by the codes that have been used to empirically measure the SFR.

An important point to keep in mind, however, is that both the
\citet{2009A&A...508..541K} and \citet{2011ApJ...737...13K} comparisons, as well as
\citet{2010A&A...512A..81F} and \citet{2011ApJ...731...62F},
illustrate that turbulent structures are resolved well only down to a scale of
10--30 grid cells; smaller scales are, to quote \citet{2009A&A...508..541K},
``significantly affected by the specific implementation of the algorithm for
solving the equations of hydrodynamics". These effects, as well as the relative
similarity with which different codes are affected, are well illustrated by
Fig.\ 3 of \citet{2011ApJ...737...13K}, which shows that a drop in compensated
velocity power spectra sets in about one order of magnitude below the Nyquist
wave number; i.e., at about 20 cells per wavelength, in essentially all codes.
Even the world's highest-resolution simulation of supersonic turbulence to
date, with $4096^3$ grid cells, only yields a rather limited inertial
range \citep{2013arXiv1306.3989F}.  The possible influence of the
magnetic Prandtl number, effectively close to unity in numerical
experiments, while typically much larger in the ISM, should also be kept in mind
\citep{2011PhRvL.107k4504F,2012PhRvE..85b6303S,2013NJPh...15a3055B,2013NJPh...15b3017S}.

In numerical experiments, the SFR is measured using
`sink particles' that can accumulate gas from their surroundings. The use of
sink particles is convenient, because one can directly measure the conversion of
gas mass into stellar mass.  It is also inevitable, since it would
otherwise be very difficult to handle the mass that undergoes gravitational
collapse. 
The sink-particle technique goes back to \citet{1982ApJ...258..270B}, who used a
single, central `sink cell' in a grid code. Since then, sink-particle methods
have been described and discussed e.g.\ by
\citet{1995MNRAS.277..362B}, 
\citet{2004ApJ...611..399K}, 
\citet{2005A&A...435..611J}, 
and by \citet{2010ApJ...713..269F}. 

\citet{1995MNRAS.277..362B} applied a sink-particle technique to the modeling
of accretion in protobinary systems, using an SPH method, with a series of
criteria for deciding when to dynamically form new sink particles.  The most
important criteria are that the density exceeds a certain, fixed threshold, and
that the total energy is negative; i.e., the gas is bound. To decide whether to
accrete onto existing sink particles they used a fixed accretion radius, within
which an SPH particle would be accreted if it was gravitationally bound, with
an angular momentum implying a circular orbit less than the accretion radius,
and with auxiliary conditions imposed to resolve ambiguous cases.

\citet{2004ApJ...611..399K} applied a similar method in the context of an AMR
grid code (Orion). To decide when to create new sink particles they used
primarily the local Jeans length, which is equivalent to a density threshold criterion. 
Sink particles were created when a Jeans
length was resolved with less than four grid cells, thus combining the sink-particle 
creation criterion with the Truelove criterion \citep{1997ApJ...489L.179T},
which prevents artificial gravitational fragmentation.  To handle the creation
of very close sink particles they merged all sink particles with distances below
a given accretion radius.  To decide how rapidly to accrete onto existing sink
particles they used a method based primarily on the Bondi-Hoyle accretion
formula, complemented by taking the effects of residual angular momentum into
account. 

\citet{2010ApJ...713..269F} implemented sink particles in a version of the
Flash code \citep{2000ApJS..131..273F,2008ASPC..385..145D}, using a similar series of criteria as in
\citet{1995MNRAS.277..362B} to guard against for example spurious creation in
shocks. They checked the resulting behavior in the AMR case against the
behavior in the original SPH case. 
They demonstrated that omitting some of the sink-creation criteria may result in
significantly overestimating the number of stars created, as well as
overestimating the SFR. 

\citet{2004ApJ...611..399K}, \citet{2010ApJ...713..269F}, and
{\it Haugb{\o}lle et al.} (2013, in preparation)  seem to agree on the fact that the 
global SFR in simulations is not as sensitive to details of the sink-particle recipes
as the number and mass distribution of sink particles. It appears
that the rate of accretion is determined at some relatively moderate density
level, well below typical values adopted as thresholds for sink-particle
creation, and that various auxiliary criteria can
result in the formation of a larger or smaller number of sinks, without
affecting the global SFR very much.

An important class of effects that are largely missing from current simulations
aimed at measuring the SFR consists of (direct and indirect) effects of stellar
feedback. Bipolar outflows, for example, both reduce the final mass of the
stars (a `direct' effect), and feed kinetic energy back into the ISM, thus
increasing the virial number and reducing the SFR (an `indirect'
effect).

As discussed elsewhere in this Chapter, both comparisons of the core mass
function with the stellar initial mass function and comparisons of numerical
simulations with observations indicate that of the order of half of the mass is
lost in bipolar outflows, i.e., leading to a core-to-star efficiency 
\mbox{$\eps\approx0.3$--$0.7$} \citep{2000ApJ...545..364M,2012ApJ...761..156F}. 
In the near future, it may become possible to model
the outflow in sufficient detail to measure the extent to which accretion is
diverted into outflows, using local zoom-in with very deep AMR simulations,
which are able to resolve the launching of outflows in the immediate
neighborhood of sink particles \citep{2012MNRAS.422..347S,2012MNRAS.423L..45P,Nordlund+IAUS299}.

The indirect effect of outflow feedback has been modelled
by adopting assumptions about the mass and momentum loading of bipolar
outflows \citep[e.g.][]{2008ApJ...687..354N,2010ApJ...709...27W,2011ApJ...740..107C}.
When attempting to quantify the importance of the feedback, it is important to adopt
realistic values of the virial number. If the virial number of the initial setup is small, the
relative importance of the feedback from outflows is increased. A number of studies of
molecular clouds have found very extended inertial ranges 
\citep[e.g.,][]{2002A&A...390..307O,2004ApJ...615L..45H,2006ApJ...653L.125P,2011ApJ...740..120R}, 
and relatively smooth velocity power spectra without bumps or features at high wavenumbers 
\citep[e.g.][]{2009ApJ...707L.153P,2009A&A...504..883B}, suggesting that local feedbacks are 
not a dominant energy source, compared to the inertial energy cascade from very large scales.

Finally, a few comments on the importance of feedback from stellar radiation (see the Chapters
by Krumholz et al. and Offner et al. in this book). 
As obvious from a number of well known Hubble images, and also from simple
back-of-the-envelope estimates, the UV radiation from new-born massive stars
can have dramatic effects on the surrounding ISM, ionizing large ISM bubbles
and cooking away the outer layers of cold, dusty molecular clouds.  Such
effects have recently started to become incorporated in models of star formation 
\citep[e.g.][]{2010ApJ...711.1017P,2011ApJ...742L...9C,2011ApJ...732...20K,2012ApJ...754...71K,2012MNRAS.419.3115B}. 
In a series of papers, \citet{2012MNRAS.424..377D,2012MNRAS.427.2852D,2013MNRAS.430..234D,2013MNRAS.431.1062D}
have explored the effect of HII regions on 
the SFR. They find that the ionization feedback reduces the SFR, by an amount
that depends on the virial parameter and total mass of the cloud. 
To derive a realistic measure of the effect of feedbacks, one needs to make 
sure that other factors that affect the SFR have realistic levels as well. For example, most 
numerical studies of the effect of massive star feedback
have neglected magnetic fields, which are known to reduce the SFR by a factor of two or three 
\citep[][and \S~4.3]{2011ApJ...730...40P,2012ApJ...761..156F} even without any stellar feedback. 
Furthermore, as real clouds are generally not isolated,
inertial driving from larger scales \citep[for example gas accretion onto a cloud, see e.g.,][]{2010A&A...520A..17K,2013MNRAS.431.3196S} may also reduce the SFR.

\bigskip
\noindent
\textbf{4.2
Setups of Numerical Experiments}
\bigskip

Besides the differences in numerical methods discussed in the previous section,
numerical studies of star formation may also differ significantly with respect to their 
boundary and initial conditions. Ideally, one would like to model the whole process 
starting from cloud formation in galaxy simulations \citep[e.g.,][see also the chapter by 
Dobbs et al.~in this book]{2006ApJ...641..878T,2006MNRAS.371.1663D} and large-scale 
colliding flows \citep{2005ApJ...633L.113H,2006ApJ...643..245V,2008A&A...486L..43H,2009MNRAS.398.1082B} 
down to the formation of individual stars. However, this is beyond the current capabilities of supercomputers.
Due to their local nature (typically $\sim 1$--10 pc), star formation simulations do not couple
the formation of prestellar cores and their collapse with the large-scale processes driving the ISM turbulence, 
such as the evolution of supernova (SN) remnants, and the formation and evolution of giant molecular clouds (GMCs). 
In setting up such simulations, one can choose either to completely ignore the effect of the larger scales
\citep[e.g.][]{2005MNRAS.359..809C,2009MNRAS.392..590B,2011MNRAS.410.2339B,2011MNRAS.413.2741G}, or to model it with external forces
\citep[e.g.][]{2000ApJ...535..887K,2001ApJ...547..280H,2011ApJ...730...40P,2012ApJ...759L..27P,2012ApJ...761..156F}. 

When driving forces are included, it is usually done with periodic boundary conditions and with a random, 
large-scale force generated in Fourier space. This artificial random force is supposed to mimic a turbulent 
cascade of energy starting at a much larger scale, but its effect may be different from that of the true driving 
forces of the ISM, such as SNs, spiral arm compression, expanding radiation fronts, winds, jets, outflows, and other mechanisms. 
The SFR in the simulations depends on whether a driving force is included or not \citep[e.g.][]{2000ApJ...535..887K,2003ApJ...585L.131V}. 
Without an external force, the turbulence decays rapidly 
\citep{1997astro.ph..6176P,1998ApJ...508L..99S,1998PhRvL..80.2754M,1999ApJ...526..279P,1999ApJ...524..169M}
and star formation proceeds at a very high rate, 
while, if the turbulence is continuously driven, a lower SFR can be maintained for an extended time. Both types 
of numerical setups have drawbacks. If external forces are neglected and star formation proceeds very rapidly, 
unphysical initial conditions may strongly affect the results; if the turbulence is driven, initial conditions are forgotten 
over time, but the results may depend on the artificial force.  

Large differences between simulations are also found in their initial conditions. These can be taken to be rather artificial,
for example with stochastic velocity fields inconsistent with supersonic turbulence apart from their power spectrum 
\citep[e.g.][]{2008MNRAS.386....3C,2009MNRAS.392..590B}, or self-consistently developed by driving the turbulence over many dynamical 
times prior to the inclusion of self-gravity \citep[e.g.][]{2009ApJ...703..131O,2011ApJ...730...40P,2012ApJ...759L..27P,2012ApJ...761..156F}.
In some studies, initial conditions are taken to follow ad hoc density profiles or velocity power spectra with the purpose 
of studying the effect of those choices \citep[e.g.][]{2005MNRAS.356.1201B,2011MNRAS.413.2741G}.

\bigskip
\noindent
\textbf{4.3
Results from Large Parameter Studies}
\bigskip

Some of the first numerical studies testing the effects of the turbulence, its driving scale and the magnetic field strength on the SFR were carried out by \citet{2000ApJ...535..887K}, \citet{2001ApJ...547..280H}, and \citet{2003ApJ...585L.131V}.
Recent high-resolution simulations by \citet{2011ApJ...730...40P}, \citet{2012ApJ...759L..27P} and \citet{2012ApJ...761..156F}, covering the largest range of molecular cloud parameters available to date, demonstrate that the SFR is primarily determined by four dimensionless parameters:
\begin{enumerate}
\item the virial parameter, $\alphavir=2 E_\mathrm{kin}/E_\mathrm{grav}$,
\item the sonic Mach number, $\mach=\sigma_v/\cs$,
\item the turbulence driving parameter, $b$ in Equations~(\ref{eq_sig_rho}) and~(\ref{eq_sig_mhd}), and
\item the plasma $\beta=\langle P_\mathrm{gas}\rangle/\langle P_\mathrm{mag}\rangle$, or the Alfv\'enic Mach number $\macha$, defined as $\macha=\sigma_v/v_\mathrm{A}=\sigma_v/(\langle B^2\rangle/\langle4\pi\rho\rangle)^{1/2}$.
\end{enumerate}

These parameter studies adopted periodic boundary conditions, an isothermal equation of state, and large-scale random driving.
The turbulence was driven without self-gravity for several dynamical times, to reach a statistical steady state.
Self-gravity was then included, and collapsing regions above a density threshold, $\rho_{\rm max}$, and satisfying other conditions (see Section~4.1),
were captured by sink particles.

\citet{2011ApJ...730...40P} carried out the first large parameter study of the SFR using the Stagger-code to run simulations on uniform numerical
grids of $500^3$ and $1000^3$ computational points. They created the sink particles only based on a density threshold, $\rho_{\rm max}= 8000 \langle\rho\rangle$,
large enough to avoid non-collapsing density fluctuations created by shocks in the turbulent flow. They explored two different values of $\mach$, two
different values of $\beta$,  and seven different values of $\alphavir$. Their simulations showed that $\sfrff$ can be reduced by the magnetic field by
approximately a factor of three, that in the non-magnetized case $\sfrff$ increases with increasing values of $\mach$, and that, both with and
without magnetic fields, $\sfrff$ decreases with increasing values of $\alphavir$, confirming the predictions of their model.

\begin{figure}[t]
\centerline{\includegraphics[width=1.0\linewidth]{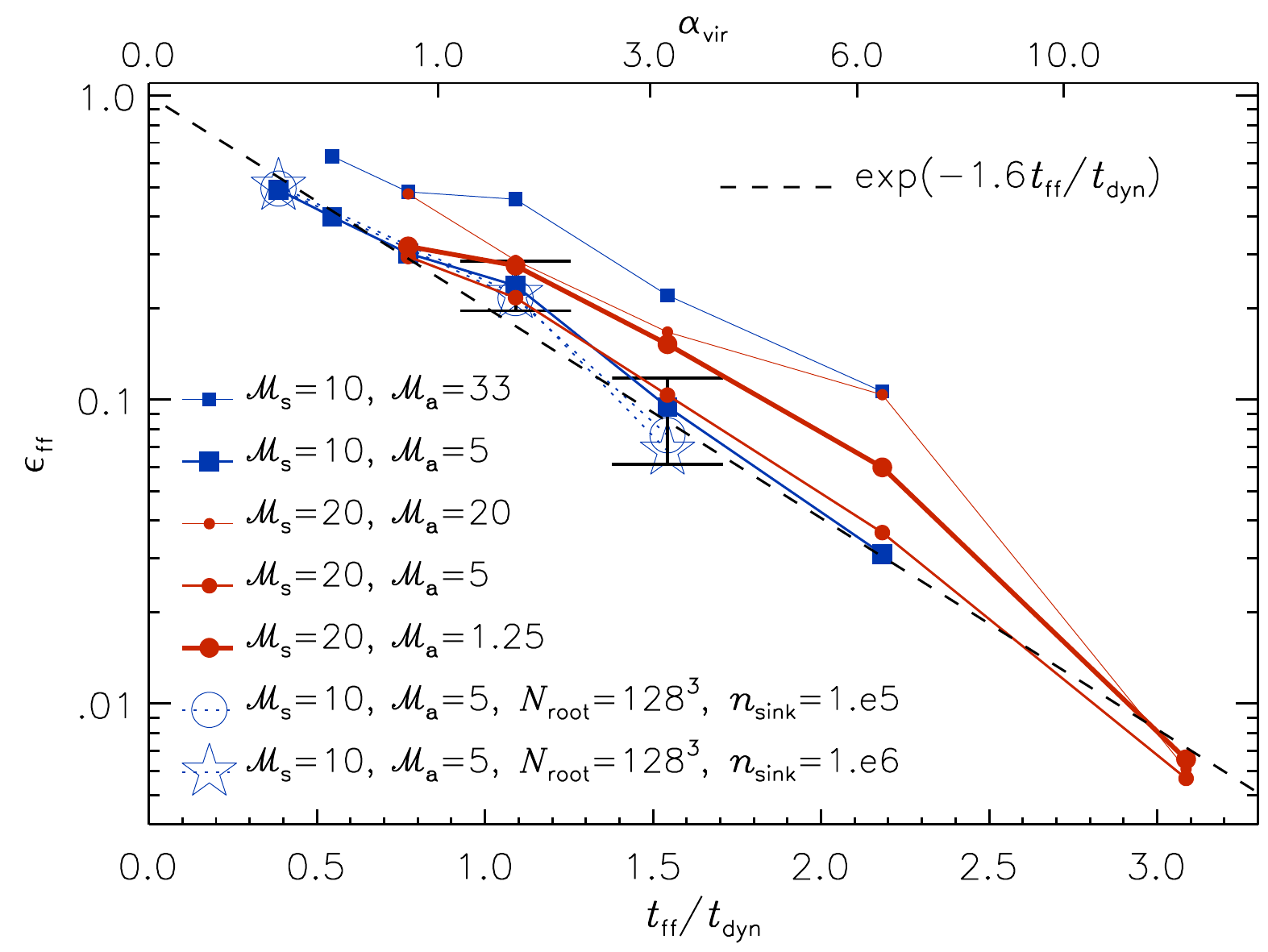}}
\caption{SFR per free-fall time, $\sfrff$, versus $t_{\rm ff}/t_{\rm dyn}$ (bottom abscissa) and $\alphavir$ (top abscissa). The symbols for each series of runs, where
only the strength of gravity is changed and $\mach$ and $\macha$ are kept constant, are connected by a line, to better distinguish
each series. The two error bars give the mean of $\sfrff$, plus and minus the rms values, for each group of five $32^3$-root-grid runs
with identical parameters ($\mach\approx 10$ and $\macha\approx 5$), but different initial conditions. The dashed line is an approximate
exponential fit to the minimum value of $\sfrff$ versus $t_{\rm ff}/t_{\rm dyn}$. From \citet{2012ApJ...759L..27P}, reproduced by permission of the AAS.
}
\label{fig:padoanSFR}
\end{figure}

In their following work, \citet{2012ApJ...759L..27P} analyzed an even larger parameter study, based on 45 AMR simulations with the Ramses code,
with a maximum resolution equivalent to $32,768^3$ computational points. Thanks to the very large dynamic range, they could adopt a threshold density
for the creation of sink particles of $\rho_{\rm max}=10^5 \langle\rho\rangle$, much larger than in the uniform-grid simulations. The creation of a sink particle
also required that the cell was at a minimum of the gravitational potential, and that the velocity divergence was negative. They explored two values of sonic 
Mach number, $\mach=10$ and 20, four values of initial Alfv\'{e}nic Mach number, $\macha=1.25$, 5, 20, 33, and seven values of the virial parameter, in the
approximate range $0.2 < \alphavir < 20$. Their results are presented in Figure~\ref{fig:padoanSFR}, and can be summarized in three points:
 i) $\sfrff$ decreases exponentially with increasing $t_{\rm ff}/t_{\rm dyn}$ ($\propto \alphavir^{1/2}$),
but is insensitive to changes in $\mach$ (in the range $10\le\mach\le20$), for constant values of $t_{\rm ff}/t_{\rm dyn}$  and
$\macha$. ii) Decreasing values of $\macha$ (increasing magnetic field strength) reduce $\sfrff$, but only to a point,
beyond which $\sfrff$ increases with a further decrease of $\macha$. iii) For values of $\macha$ characteristic of
star-forming regions, $\sfrff$ varies with $\macha$ by less than a factor of two. Therefore, \citet{2012ApJ...759L..27P}
proposed a simple law for the SFR depending only on $t_{\rm ff}/t_{\rm dyn}$, based on the empirical fit to the minimum $\sfrff$:
$\sfrff \approx \epsilon \exp(-1.6 \,t_{\rm ff}/t_{\rm dyn})$ (dashed line in Figure~\ref{fig:padoanSFR}),
where $\epsilon$ is the core-to-star formation efficiency.

\begin{figure*}[t]
\centerline{
\def\arraystretch{1.0}
\setlength{\tabcolsep}{3pt}
\begin{tabular}{cc}
\includegraphics[width=0.49\linewidth]{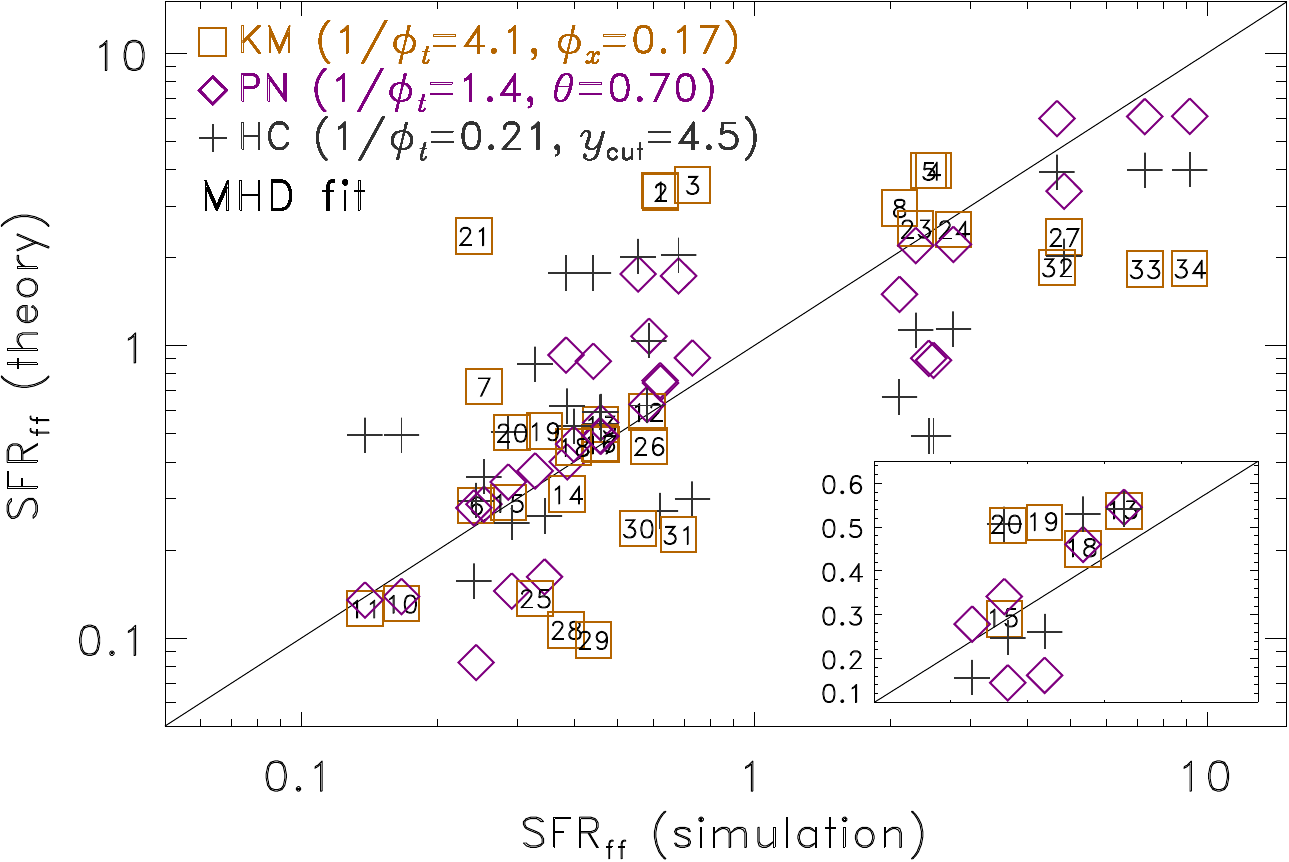} &
\includegraphics[width=0.49\linewidth]{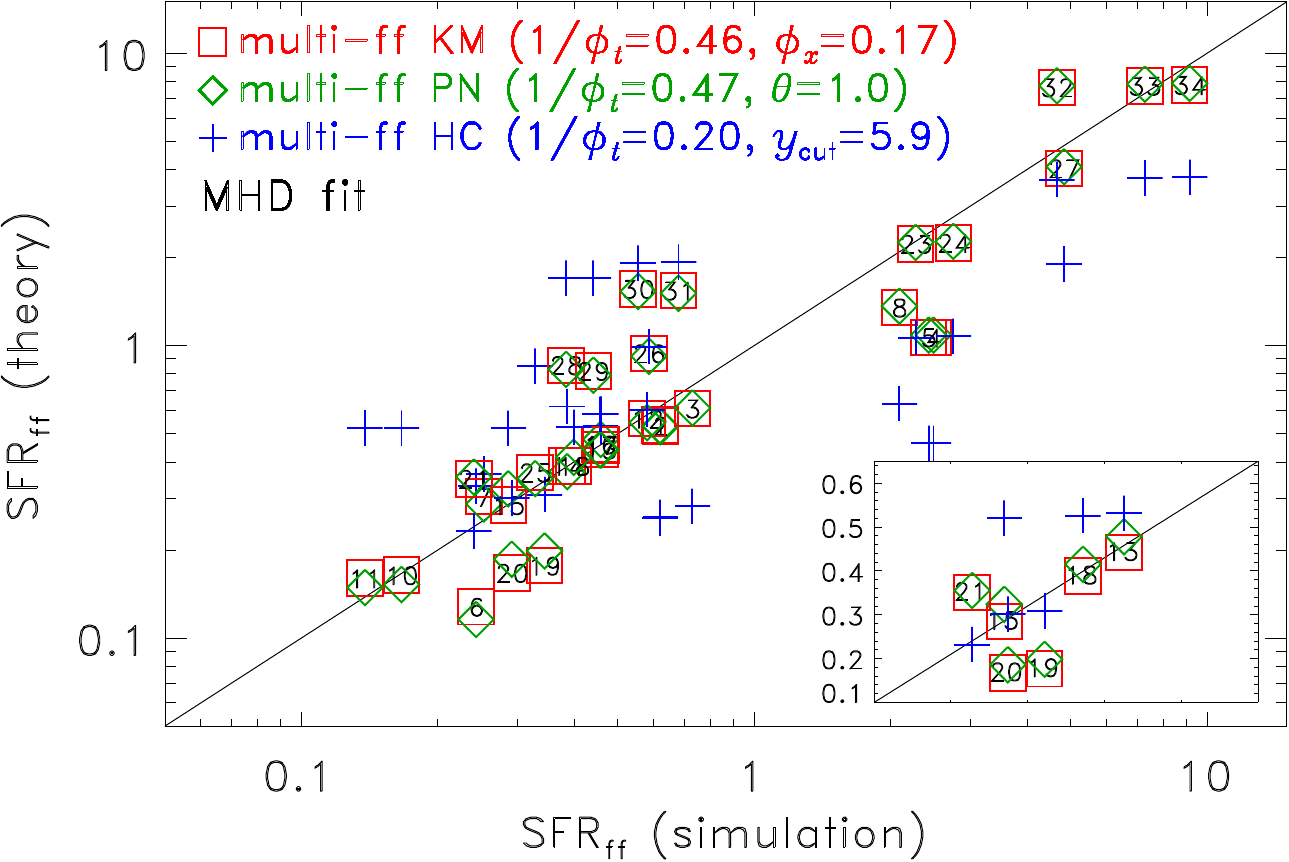}
\end{tabular}
}
\caption{Comparison of $\sfrff$ (theory) with $\sfrff$ (simulation). The left panel shows the original 
KM (boxes), PN (diamonds), and HC (crosses) theories, while the right panel shows the multi-freefall 
version of each theory defined in \citet{2011ApJ...743L..29H}. The multi-freefall prescription is superior 
to all single-freefall models and provides good fits to the numerical simulations (the insets show blow-ups of the MHD simulations; the x-range of the insets is identical to the y-range). The multi-ff KM and multi-ff PN models agree to within a factor of three with any 
of the numerical simulations over the two orders of magnitude in SFRs tested. The simulation number in Table~2 in \citet{2012ApJ...761..156F} is given in the boxes for each $\sfrff(\mathrm{simulation})$. From \citet{2012ApJ...761..156F}, reproduced by permission of the AAS.}
\label{fig:model_comparison}
\end{figure*}

These results were confirmed and extended in a third parameter study by \citet{2012ApJ...761..156F}, based on 34 uniform grid simulations with the Flash code, using a resolution of up to 512$^3$ computational points (plus one AMR run with maximum resolution equivalent to $1,024^3$ computational points).
Six of their runs include a magnetic field, covering the range $1.3 \le \macha\le 13$, but all with approximately the same sonic Mach number,
$\mach\approx 10$, so the lack of dependence of $\sfrff$ on $\mach$ found by \citet{2012ApJ...759L..27P} could not be
verified. On the other hand, the 28 runs without magnetic fields span a wide range of values of sonic Mach number, $2.9 \le \mach \le 52$,
which allowed \citet{2012ApJ...761..156F} to confirm the analytical and numerical result of \citet{2011ApJ...730...40P}, that in the non-magnetized
case $\sfrff$ increases with increasing $\mach$. \citet{2012ApJ...761..156F} also studied the effect of varying $b$
(the ratio of compressible to total energy in the turbulence driving), and carried out a systematic comparison of their simulations
with the predictions of SFR models and observations (see Sections~4.4 and 5.1, respectively).

\citet{2012ApJ...761..156F} found that both $\mach$ and $b$ can introduce order-of-magnitude variations in $\sfrff$, in the absence
of magnetic fields. Increasing $b$ and $\mach$ produces a wider density PDF (see eqs.~\ref{eq_sig_rho} and~\ref{eq_sig_mhd}) 
and thus pushes a larger fraction of gas above the critical density for star formation, increasing $\sfrff$ (a larger $\mach$ results in 
a larger critical density, but also in a shorter free-fall time at such density).
For purely compressive driving, \citet{2012ApJ...761..156F} found a $4\times$ higher $\sfrff$, when
increasing $\mach$ from 5 to 50. For fixed $\mach=10$, which is a reasonable average Mach number for Milky Way clouds,
they found that purely compressive (curl-free, $b=1$) driving yields about $10\times$ higher $\sfrff$ compared to purely solenoidal
(divergence-free, $b=1/3$) driving. 
The increase of $\sfrff$ for compressive driving is caused by the denser structures (filaments and cores) produced with such driving, which are more
gravitationally bound than the structures produced with purely solenoidal driving. 

Increasing the magnetic field strength and thus decreasing $\beta$, reduces $\sfrff$.
Numerical simulations by \citet{2011ApJ...730...40P}, \citet{2012ApJ...759L..27P} and by \citet{2012ApJ...761..156F} quantify
the effect of the magnetic field and find a maximum reduction of the SFR by a factor of 2--3 in strongly magnetized, trans- to
sub-Alfv\'enic turbulence compared to purely hydrodynamic turbulence. This is a relatively small change in $\sfrff$ compared
to changes induced by $\alphavir$, $\mach$, and $b$, in the absence of magnetic fields.

\bigskip
\noindent
\textbf{4.4
Comparison with Theoretical Models}
\bigskip

To compare them with numerical results, we separate the theories of the SFR into six cases (see Section~3.3), 
named `KM', `PN', `HC', and `multi-ff KM', `multi-ff PN', `multi-ff HC', following the notation in \citet{2011ApJ...743L..29H} and \citet{2012ApJ...761..156F}. 
The first three represent the original analytical derivations by \citet{2005ApJ...630..250K}, \citet{2011ApJ...730...40P}, and \citet{2011ApJ...743L..29H}, 
while the last set of three are all based on the multi-freefall expression of eq.~(\ref{sfr_int}). Notice that the HC models, here and in \citet{2012ApJ...761..156F}, 
refer to the aforementioned {\it simplified} version of the HC model without a scale dependence in the density variance, which yields eq. (\ref{eq:sfrff_basicsolution}). 
In case of the {\it complete} theory, the integral of
eq. (\ref{sfr_int}) can no longer be performed analytically. Besides this important point, the difference among the multi-freefall models is the expression 
for the critical density (the lower limit of the integral in eq.~\ref{sfr_int}). Table~\ref{tab:theories} summarizes the differences among all six theories 
following \citet{2012ApJ...761..156F}, who also extended all theoretical models to include the effects of magnetic pressure (except PN, where this 
dependence was already included).

The comparison between the theoretical and the numerical $\sfrff$ is shown in Figure~\ref{fig:model_comparison} 
(left panels: KM, PN, HC; right panels: multi-freefall KM, PN, HC). Any of the multi-freefall theories 
provides significantly better fits to the numerical simulations than the single-freefall models 
($\chisqred=1.3$ and $1.2$ for multi-ff KM and PN; $\chisqred=5.7$ and $1.8$ 
for KM and PN). Note that HC and multi-ff HC are both multi-ff models, but \citet{2011ApJ...743L..29H} defined them this 
way to distinguish their model where the critical density includes both thermal and turbulent support (called HC) from the one where
only thermal support is included (called multi-ff HC) (see Table~\ref{tab:theories} and \S~3.2).

The simplified HC model, although originally designed as a multi-freefall model, does not fit the simulations as well as the multi-ff KM and multi-ff PN models. 
This is likely related to the definition of the critical density in HC, where it scales with $\mach^{-2}$, 
while in KM and PN it scales with $\mach^2$ 
(see Section~3.2, Table~\ref{tab:theories}, and the derivation in \citet{2012ApJ...761..156F}).
A caveat of the present comparison is that the density PDF in the models is assumed to be perfectly log-normal. Accounting for deviations of
the PDF in future semi-analytical models may further improve the agreement with the simulations. 

Figure~\ref{fig:model_comparison} shows that the multi-freefall KM and PN theories provide 
good fits to the numerical simulations. However, some data points in Figure~\ref{fig:model_comparison} 
deviate from the theoretical predictions by a factor of 2--3. The simulations that deviate the most 
have relatively low numerical resolution. A careful resolution study 
of these models shows that they converge to the diagonal line in Figure~\ref{fig:model_comparison}. 
Notably, simulation models 32, 33, and 34 with numerical resolutions of $256^3$, $512^3$, and 
$1024^3$ grid cells (in the upper right corner of Figure~\ref{fig:model_comparison}) clearly converge 
to the diagonal line with increasing resolution. The same is true for the MHD models 19 and 20 with 
$256^3$ and $512^3$ resolution in the inset of Figure~\ref{fig:model_comparison}. This convergence 
with increasing resolution supports the conclusion that the multi-freefall theories for the SFR provide a 
good match to numerical simulations where the interplay of self-gravity and supersonic turbulence is 
the primary process controlling the SFR.

\section{\textbf{THEORY AND SIMULATIONS VERSUS OBSERVATIONS}}

\noindent
\textbf{5.1
Model Predictions versus Observed SFR}
\bigskip

\begin{figure}[t]
\centerline{\includegraphics[width=1.0\linewidth]{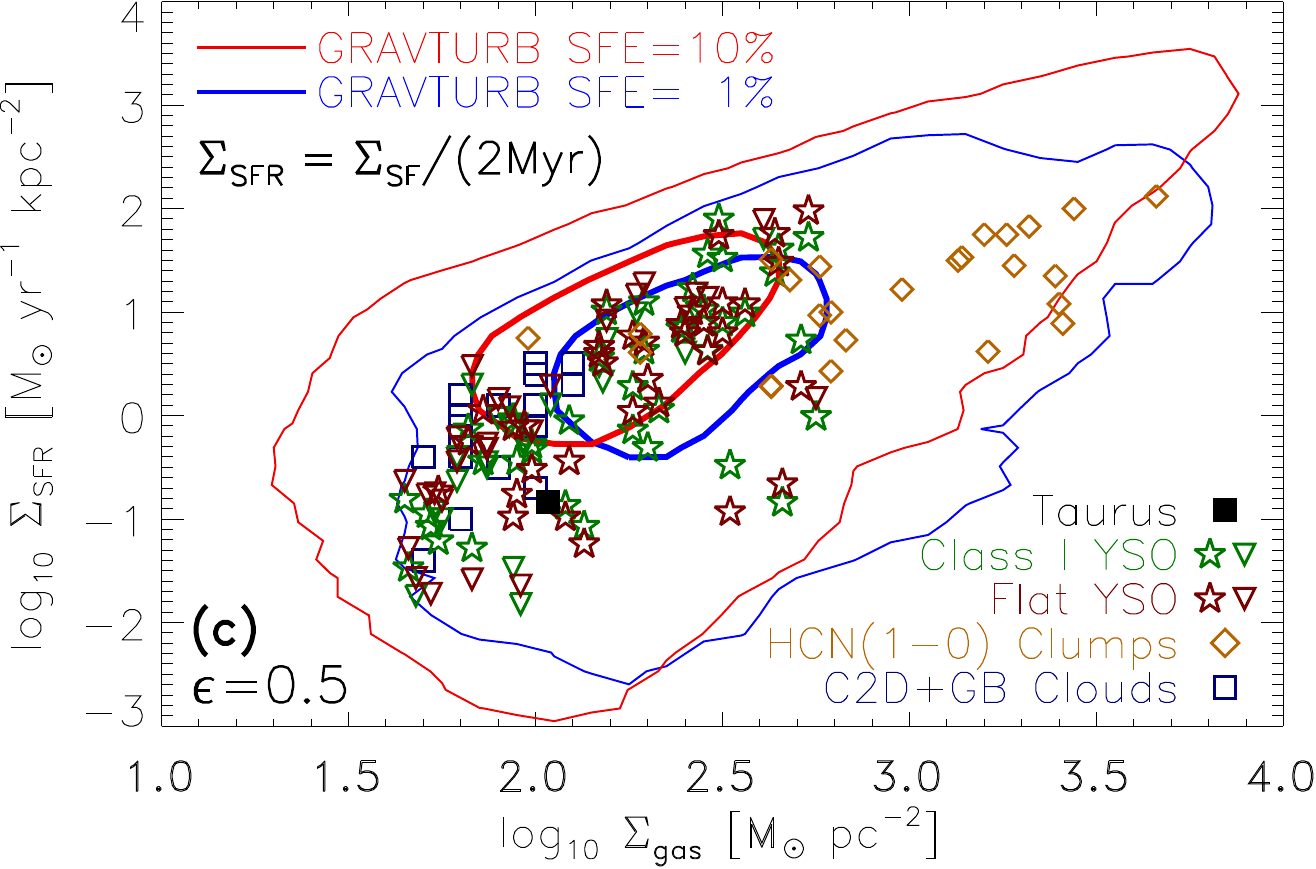}}
\caption{Star formation rate column density $\sigsfr$ versus gas column density $\siggas$ for Milky Way 
clouds compiled in \citet{2010ApJ...723.1019H} (symbols) and in the GRAVTURB simulations by \cite{2012ApJ...761..156F} (contours). 
Individual data points are Taurus: filled black box \citep[data from][]{2008ApJ...680..428G,2010ApJ...721..686P,2010ApJS..186..259R}, 
Class I YSOs and Flat YSOs: green and red stars and upper-limits shown as downward-pointing triangles, HCN(1--0) Clumps: 
golden diamonds \citep[data from][]{2004ApJ...606..271G,2005ApJ...635L.173W,2010ApJS..188..313W}, and C2D+GB Clouds: 
dark blue boxes \citep[data from][]{2009ApJS..181..321E}. Blue and red contours show data from numerical simulations by 
\citet{2012ApJ...761..156F} for a star formation efficiency $\sfe=1\%$ (blue) and $\sfe=10\%$ (red). The thick contours enclose 
50\% of all ($\siggas,\,\sigsfr$) simulation pairs, while the thin contours enclose 99\%. All simulation data were scaled to a local 
core-to-star efficiency of $\epsilon=0.5$ \citep{2000ApJ...545..364M}, providing the best fit to the observational data.
From \citet{2012ApJ...761..156F}, reproduced by permission of the AAS.
}
\label{fig:sfrcoldens}
\end{figure}

A comparison of simulations from \citet{2012ApJ...761..156F} with observations of $\sigsfr$ 
in Milky Way clouds by \citet{2010ApJ...723.1019H} is shown in Figure~\ref{fig:sfrcoldens}. 
The observational data are from Galactic observations of clouds and young stellar objects 
(YSOs) identified in the Spitzer Cores-to-Disks (c2d) and Gould's-Belt (GB) surveys 
\citep{2009ApJS..181..321E}, of massive dense clumps 
\citep{2004ApJ...606..271G,2005ApJ...635L.173W,2010ApJS..188..313W}, and of the
Taurus molecular cloud \citep{2008ApJ...680..428G,2010ApJ...721..686P,2010ApJS..186..259R}. 
Simulation data are shown as contours for $\sfe=1\%$ (blue) and $\sfe=10\%$ (red).

The simulation data from \citet{2012ApJ...761..156F} shown as contours in Figure~\ref{fig:sfrcoldens} were transformed into the observational space by measuring $\siggas$ and $\sigsfr$ with a method as close as possible to what observers do to infer $\sigsfr$-to-$\siggas$ relations \citep[see, e.g.,][]{2008AJ....136.2846B,2010ApJ...723.1019H}, including the effects of telescope beam smoothing. This was done by computing two-dimensional projections of the gas column density, $\siggas$, and the sink particle column density, $\sigsf$, along each coordinate axis: $x$, $y$, $z$. Each of these 2D maps was smoothed to a resolution $(N_\mathrm{res}/8)^2$ for a given 3D numerical resolution \mbox{$N_\mathrm{res}^3=128^3$--$1024^3$ grid cells} \citep[for a complete list of simulations and their parameters, see Table~2 in][]{2012ApJ...761..156F}, such that the size of each pixel in the smoothed maps slightly exceeds the sink particle diameter (which is 5 grid cells). We then search for pixels with a sink particle column density greater than zero, $\sigsf>0$, and extract the corresponding pixel in the gas column density map, which gives $\siggas$ in units of $\msol\,\pc^{-2}$ for that pixel. The SFR column density is computed by taking the sink particle column density $\sigsf$ of the same pixel and dividing it by a characteristic timescale for star formation, $\tsf$, which yields $\sigsfr=\sigsf/\tsf$ in units of $\msol\,\yr^{-1}\,\mathrm{kpc}^{-2}$. The simplest choice for $\tsf$ is a fixed star formation time, $\tsf=2\,\mathrm{M}\yr$, based on an estimate of the elapsed time between star formation and the end of the Class II phase \citep[e.g.,][]{2009ApJS..181..321E,2010ApJ...722..971C}. This is also the $\tsf$ adopted by \citet{2010ApJ...724..687L} and \citet{2010ApJ...723.1019H} to convert young stellar object (YSO) counts into an SFR column density (see Section~2.4), so we used it here as the standard approach. However, \citet{2012ApJ...761..156F} also experimented with two other choices for $\tsf$, which gave similar results, considering the large dispersion of the observational data.

The \citet{2010ApJ...723.1019H} sample of SFR column densities for Galactic clouds in 
Figure~\ref{fig:sfrcoldens} covers different evolutionary stages of the clouds, such that a 
single $\sfe$ for the whole sample is unlikely. However, we can reasonably assume 
SFEs in the range 1\% to 10\% \citep[][]{2009ApJS..181..321E,2013ApJ...763...51F}. 
For a typical $\sfe=3\%$, \citet{2012ApJ...761..156F} find a best-fit match of their 
simulations to the \citet{2010ApJ...723.1019H} sample by scaling the simulations 
to a core-to-star efficiency $\epsilon=0.5$, in agreement with theoretical 
results \citep{2000ApJ...545..364M}, with observations 
of jets and outflows \citep[][]{2002A&A...383..892B}, and with independent numerical 
simulations that concentrate on the collapse of individual cloud cores \citep[e.g.,][]{2010ApJ...709...27W,2012MNRAS.422..347S}.

\begin{figure*}[t]
\centerline{
\includegraphics[width=0.5\linewidth]{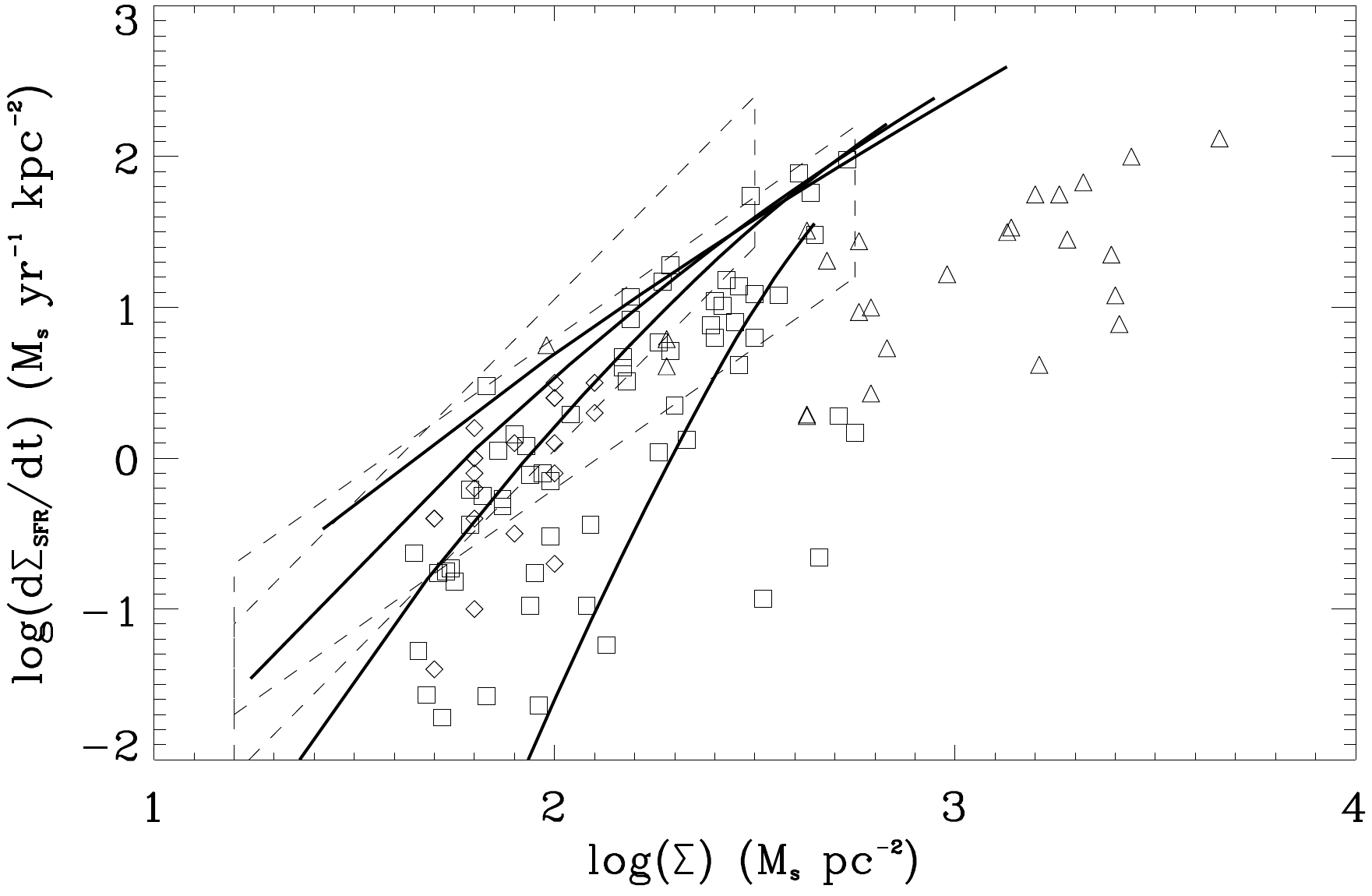}
\includegraphics[width=0.5\linewidth]{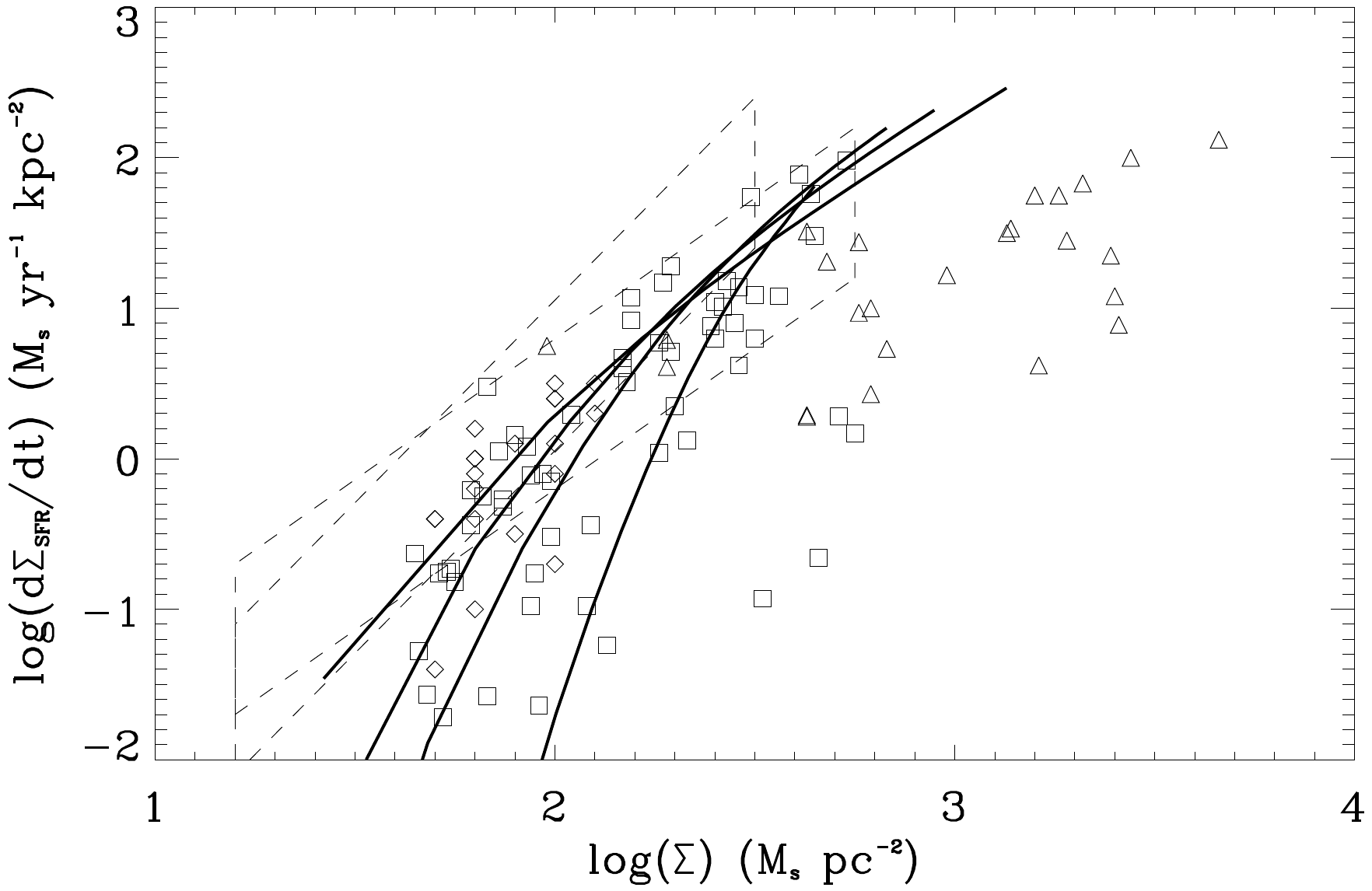}
}
\caption{Star formation rate per unit area, $\sigsfr$, as a function of gas surface density, $\siggas$. Solid lines: results obtained with the {\it complete} HC SFR theory \citep{2013ApJ...770..150H} for cloud sizes $R_c=20$, 5, 2 and 0.5 pc, from left to right, in the case of isothermal (left panel) and non-isothermal (right panel) gas, for $y_{cut}=0.25$ and $b=0.5$, and assuming a density--size relation, $\rho\propto R_c^{-0.7}$. Each curve is obtained by varying the normalization of the density--size relation (see \citet{2013ApJ...770..150H} for details). The data correspond to observational determinations by \citet{2010ApJ...723.1019H} for massive clumps (triangles) and 
molecular cloud YSOs (diamonds+squares), and by \citet{2011ApJ...739...84G}
(bracketed areas). From \citet{2013ApJ...770..150H}, reproduced by permission of the AAS.
}
\label{fig_sfr_HC}
\end{figure*}

The simulation data in Figure~\ref{fig:sfrcoldens} are also consistent with the 
Milky Way cloud samples in \citet{2010ApJ...724..687L} and \citet{2011ApJ...739...84G}, 
showing that $\sigsfr$ can vary by more than an order of magnitude for any given $\siggas$. 
These variations can be explained by variations in $\alphavir$
\citep{2011ApJ...743L..29H,2012ApJ...759L..27P}, $\mach$ 
\citep{2012ApJ...760L..16R,2013arXiv1307.1467F}, $b$ 
\citep{2012ApJ...761..156F}, and $\beta$ \citep{2011ApJ...730...40P,2012ApJ...761..156F} from cloud to cloud. 
Besides these physical effects primarily related to the statistics of self-gravitating turbulence, the $\sigsfr$--$\siggas$ 
relation is also somewhat affected by observational issues \citep[e.g., telescope resolution; see][]{2010ApJ...723.1019H,2012ApJ...761..156F} 
and by variations in the metallicity \citep{2012ApJ...745...69K}. The latter effect introduces uncertainties by a factor of about two \citep{2012MNRAS.426..377G}.

Figure \ref{fig_sfr_HC} compares the SFR obtained with the {\it complete} HC formalism
for four different cloud sizes $R_c=0.5$, 2, 5 and 20 pc, with the data from \citet{2010ApJ...723.1019H} and 
\citet{2011ApJ...739...84G}, for the case of both isothermal and non-isothermal gas, for $y_{cut}=0.25$ and $b=0.5$ 
(approximate equipartition between solenoidal and compressive modes of turbulence). For such typical Milky Way molecular 
cloud conditions, the agreement is fairly reasonable. The observed scatter, with no one-to-one correspondence between the SFR 
and the surface density, is well explained by the strong dependence of the SFR upon the cloud's size/mass, besides the effect of 
the turbulence parameters. The theory also 
adequately reproduces the observational SFR values of \citet{2011ApJ...739...84G} (bracketed areas) for the adequate (large) 
cloud sizes, down to the very low density regime, $\Sigma\sim 30$ M$_\odot$ pc$^{-2}$. This supports the suggestion that there 
is no real step-function threshold condition for star formation and that the latter can occur even in rather low-density regions, although
at a much lower rate.

In Figure~\ref{fig_sfr_HC}, the HC model is applied assuming a density--size relation, $\rho\propto R_c^{-0.7}$, which determines the
dependence on $\siggas$ seen in the theoretical curves. This density--size relation is confirmed by observations where clouds 
are defined by constant $^{13}$CO brightness contours \citep{2010ApJ...723..492R}. However, when clouds are defined by constant surface density contours, one 
gets a different density--size relation, $\rho\propto R_c^{-1.0}$ \citep{2010A&A...519L...7L}, which changes the position of the theoretical 
curves of Figure~\ref{fig_sfr_HC} slightly (the curves would be a bit more vertical and closer to one another). This highlights the importance of adopting unambiguous 
definitions of MCs in both theoretical models and observational works, or rather of deriving theoretical models that do not depend on MC definitions.

\bigskip
\noindent
\textbf{5.2
Density PDF in MCs}
\bigskip

Several studies have tried to compare the PDF predicted by supersonic turbulence with measurements in MCs. 
The pdf of a MC was first shown to be consistent with a log-normal function by \citet{1997ApJ...474..730P}. They used
the extinction measurements of IC 5146 by \citet{1994ApJ...429..694L}, specifically their $\sigma_{A_{\rm V}}-A_{\rm V}$
relation, to constrain the shape and standard deviation of the three-dimensional density PDF, and the density power spectrum.
They concluded that the density PDF was consistent with a log-normal, that the standard deviation was $\sigma_{\widetilde{\rho}}=5.0\pm0.5$,
that the slope of the density power spectrum was $-2.6\pm0.5$, and that density fluctuations had to reach at least a scale of 0.03 pc. 
Given the Mach number derived for IC 5146, $\calm_{\rm s}\approx 10$, these results were all in good agreement with their numerical 
simulations of super-Alfv\'{e}nic turbulence, showing that $\sigma_{\widetilde{\rho}}=b \calm_{\rm s}$, with $b\approx 0.5$.

\citet{2010MNRAS.403.1507B} derived analytically the relation between the variance of the three-dimensional density field and the 
observed variance and power spectrum of the projected density. Applied to the Taurus MC, this method gave a density variance 
yielding $b=0.48^{+0.15}_{-0.11}$, similar to the case of IC 5146, and consistent with that of turbulence simulations driven with a mixture 
of solenoidal and compressional modes \citep{2010A&A...513A..67B,2011ApJ...727L..21P}. 

The relation between the Mach numbers and the variance of the projected density can be derived directly from simulations.
\citet{2012ApJ...755L..19B} found that the variance, $\sigma_S$, of the logarithm of the column density, $S=\rm{ln}(\Sigma/\Sigma_0)$ (where
$\Sigma_0$ is the mean column density), depends on the sonic Mach 
number, and follows the relation $\sigma_S^2 = A\,{\rm ln}\left[1+b^2 \calm_{\rm s}^2\right]$, where $A\approx 0.11$, for a range of values 
of Mach numbers, $0.4\le \calm_s \le 8.8$ and $0.3\le \calm_{\rm A} \le 3.2$. They also used that relation to map the Mach number in the 
Small Magellanic Cloud \citep{2010ApJ...708.1204B}. By comparing with eq.~(\ref{eq_sig_rho}), one gets
that the ratio of density and column density variances is independent of the Mach number, $\sigma_S^2/\sigma_s^2=A$. 
In a subsequent paper, \citet{2013ApJ...771..122B} studied the role of radiative transfer effects on the derivation of the column density variance 
using molecular emission lines. They showed that the parameter $A$ is quite sensitive to the optical depth of the observed line, so
extinction maps, rather than molecular emission lines, are the method of choice to directly constrain the density PDF, unless opacities and
molecular abundances are known. Similar conclusions about the uncertainties in deriving column density PDFs in MCs with CO emission lines 
were also reached by \citet{2009ApJ...692...91G}, through a comparison of column density measured with dust emission and extinction. 
They also found that the relation between the mean and the variance of the projected density were consistent with a log-normal distribution.

Besides deriving the density variance, a more ambitious goal is to derive the whole PDF using MC observations.  
This was first done by \citet{1999ApJ...525..318P}, who compared a simulation of supersonic and super-Alfv\'{e}nic turbulence
with a $^{13}$CO survey of the Perseus region. Radiative transfer effects were properly accounted for, because the comparison 
was done using synthetic observations of the simulation, generated with a three-dimensional non-LTE radiative transfer code
\citep{1998ApJ...504..300P}. However, besides radiative transfer, the interpretation of molecular emission lines would also require 
a careful study of molecular abundances, in order to probe a more extended density range. The interpretation of extinction maps
is more straightforward, at least for MCs that are far enough from the galactic plane to avoid confusion,
and close enough to guarantee a large number of background stars. Moreover, \citet{2010MNRAS.405L..56B} developed a statistical method to reconstruct the volumetric (3D) density PDF from observations of the column (2D) density PDF and power spectrum.

Extinction maps were used by \citet{2009A&A...508L..35K} to derive column density PDFs of many nearby clouds. 
They found that the PDFs are consistent with a log-normal function at all densities in non-star-forming clouds. In regions of active star formation,
the PDF is log-normal at intermediate and low densities, and a power law at high densities, consistent with the results of supersonic
turbulence with self-gravity. Other studies based on extinction maps used the cumulative PDF of column density, and found again deviations
from the log-normal at high extinctions in regions of active star formation \citep{2010MNRAS.406.1350F,2013A&A...553L...8K}.
The relation between extinction-based cumulative PDF of column density and the SFR was derived even more explicitly by 
\citet{2010ApJ...724..687L}, who found a linear correlation between the SFR and the mass of gas with extinction $A_{K}>0.8$~mag.

Recently, FIR dust emission maps from Herschel have also been used to derive the PDF of column density in MCs. For example,
\citet{2012A&A...540L..11S,2013ApJ...766L..17S} have shown that PDFs from different regions of the Rosette MC deviate from the log-normal, although 
the PDF tails are not well described by single power laws. Cumulative column density PDFs from Herschel maps are found to agree
with cumulative PDFs from near- and mid-infrared extinction maps, and are therefore a powerful probe of the structure of infrared dark clouds 
\citep{2013arXiv1305.6383K}.

\bigskip
\noindent
\textbf{5.3
Star Formation Thresholds in MCs}
\bigskip

As already mentioned in Section~2.5, studies of nearby clouds, based on a variety of methods (near-IR dust extinction, sub-mm dust emission, or 
molecular emission lines) have demonstrated the existence of a column density threshold for star formation of approximately 
120~$M_\odot$~pc$^{-2}$, below which star formation appears to be rare. This threshold should not be interpreted as a strict step-function condition for star
formation (presumably, the SFE continues to increase at larger $\sigmagas$, up to a maximum equal to $\epsilon$, the core-to-star efficiency, 
when the characteristic $\sigmagas$ of cores is reached --see Section~2.5). 
The idea of the threshold is to {\it exclude} the large mass of the cloud that is currently not forming stars. Understanding how to 
stop this mass from eventually forming stars lies at the heart of the SFE problem for clouds and galaxies.
The correlation between the mass fraction of gas with surface density above the threshold and the SFR/area found by
\citet{2010ApJ...723.1019H} and \citet{2010ApJ...724..687L} may become even stronger at even larger column densities \citep{2011ApJ...739...84G}. 
This is suggested for example by the increase in the SFE toward smaller (denser) scales \citep{2013ApJ...763...51F}, with the mass in dense cores being 
comparable to the mass in YSOs \citep{2008ApJ...684.1240E,2008ApJ...683..822J}. 

It is nevertheless interesting to contrast the observed threshold with the critical volume 
density of the SFR models, $\rho_{\rm crit}$. Indeed, \citet{2010ApJ...724..687L} have argued that the column density threshold they found could 
also be interpreted as a volume density threshold of approximately $10^4$~cm$^{-3}$, and the equivalent mean volume density within the column density
threshold for the c2d+GB clouds is about 6000~$\cmv$ ({\it Evans et al.}, in preparation). The mass per unit length threshold for star formation found 
by \citet{2010A&A...518L.102A} in MC filaments probed by Herschel can also be viewed both as a column or volume density threshold, because the 
filaments they select have a well defined characteristic thickness, independent of their surface density \citep{2011A&A...529L...6A,2013A&A...553A.119A}.

We have demonstrated in Section~3.2 that the general expression for the theoretical critical density, eq.~(\ref{rho_crit}), as well as the equivalent expressions
from the KM model, eq.~(\ref{rho_crit_KM}) and the PN model, eq.~(\ref{rho_crit_PN}), yield a constant value, $n_{\rm H,\, crit}=3.0\times 10^4$~cm$^{-3}$  
(see eq.~\ref{rho_crit_const}). This is not true for the HC model, where the critical density, given by eq.~(\ref{rho_crit_HC}), depends on the cloud size (the larger
the cloud the smaller the critical density), even after assuming a standard linewidth-size relation. We conclude that, in the case of the KM and PN models, 
one would expect observational signatures of a star formation threshold approximately independent of cloud properties, at least within a MC sample 
following approximately the same velocity-size relation. Based on the HC model, instead, one would expect the threshold to be approximately the same 
only for MCs of the same size, and to increase toward smaller cloud sizes.

The observed threshold of  \citet{2010ApJ...724..687L}, $10^4$~cm$^{-3}$, is a few times smaller than the above value of $n_{\rm H,\, crit}$.
However, as mentioned above, it is not a strict step-function condition for star formation, and it is to be expected that the correlation
between the SFR and the gas mass fraction above the threshold would become increasingly stronger as the observed threshold is increased toward
the theoretical value. If the theoretical value were reached, then the SFR per unit mass should be approximately the mass fraction above 
$n_{\rm H,\, crit}$ divided by the free-fall time of $n_{\rm H,\, crit}$, multiplied by the local core-to-star efficiency. In other words, 
most of the gas denser than  $n_{\rm H,\, crit}$ should be collapsing or about to start to do so.

The prediction for high-mass star-forming cores would probably be different than for local MCs. The cores in the samples of 
\citet{1997ApJ...476..730P} and \citet{2003ApJS..149..375S}, for example, have rms velocities well in 
excess of the Larson relations for nearby clouds. Assuming that their virial parameter is not very different from that of nearby clouds, 
their larger rms Mach number at a given size (their warmer temperature probably does not compensate the increased linewidth) 
would imply a larger value of $n_{\rm H,\, crit}$ than in nearby clouds. On the other hand, if the HC model were correct, the
threshold density in these regions should not increase with the Mach number.

\section{\textbf{CONCLUSIONS AND FUTURE DIRECTIONS}}

\noindent
\textbf{6.1
Summary and Conclusions}
\bigskip

Progress on theory, simulations, and observations of star formation has been substantial since Protostars \& Planets V. We have summarized this progress, focusing on the observational and theoretical estimates of the SFR in individual clouds, with some reference to larger-scale star formation on galaxy scales.

Observationally (Section~2) we now have extinction maps of nearby clouds, defining cloud structures, surface densities, and masses, with far less uncertainty than was inherent in molecular line maps. Most of the cloud mass lies at low surface densities, well below 100 \msunpc. The surface densities follow a log-normal PDF at low extinctions ($\av <$ few mag), but clouds with active star formation show power law tails. Maps of dust continuum emission show that the denser parts of clouds are highly filamentary, with dense cores lying along filaments containing only a very small fraction of the cloud mass. 
We also have quite complete censuses of YSOs in all the larger nearby ($d < 500$ pc) molecular clouds, allowing robust estimates of SFRs averaged over the half-life of infrared excesses ($\sim 2$ Myr). These are quite low compared to what would be expected if clouds were collapsing at the free-fall rate, calculated from the mean cloud density, and making stars with unity efficiency; $\sfrff \sim 0.01$ for the clouds as a whole, but the scatter is large. If attention is focused on regions above a surface density ``threshold" of about 120 \msunpc, the dispersion in SFR per mass of this ``dense" gas is much smaller, essentially consistent with observational uncertainties, and the mean $\sfrff$ doubles to $\sim 0.02$. Thus, most of the mass of nearby clouds is quite inactive in star formation.  The mean depletion time of the nearby clouds is about 140 Myr; if only 10\% of the mass forms clouds that form stars before dissipating, the galaxy scale depletion time of about 1--2 Gyr can result from these characteristic values of local efficiency: star formation may thus be seen as the ultimate outcome of an `inefficient hierarchy'.

The fact that star formation is slow compared to free-fall times, and ultimately inefficient (mass depletion times are much greater than structure lifetimes) may be understood from current theories (Section~3) as essentially a consequence of an excess of the turbulent kinetic plus magnetic energies over the gravitational energy of a cloud.
This understanding has come about as a consequence of numerical efforts that resulted in large parameter studies based on many simulations (Section~4). A number of formalisms addressing the problem of estimating the SFR in clouds with magnetized turbulence have been developed. Their predictions have been checked against both idealized simulations (Section~4) of such clouds, and against observations (Section~5). While formulations differ in detail, the differences have been narrowed down and clarified in recent papers, as discussed in this review.

As indicated by the theoretical models (Section~3), and illustrated in practice by numerical simulations (Section~4), in addition to the virial ratio there are a number of other factors that also influence the SFR; among them the sonic and Alfv{\'e}nic Mach numbers, and the relative importance of compressive and incompressive components of the random forces driving the turbulence.
These parameters are all included in the current theoretical models for the SFR, and the best theories can successfully predict the SFR within a factor of 2--3 (Figure~\ref{fig:model_comparison}), over at least two orders of magnitude in SFR and for a wide range of parameter space, which is very encouraging, given the simplified assumptions in the analytical theories (most importantly, the current assumption of a purely log-normal PDF together with a volume-density threshold for star formation).
The importance of radiative feedback from early type stars is also broadly appreciated and accepted, as is the possible importance of kinetic feedback from low-mass stars (see the Chapter by Krumholz et al.~in this book). These latter aspects are, however, very difficult to incorporate into the analytical theories, and are also non-trivial and computationally costly to include in numerical simulations. We thus expect further developments of both analytical and numerical models addressing these issues in the near future

\bigskip
\noindent
\textbf{6.2
Future Directions}
\bigskip

An important future step will be to apply what we have learned here about the SFR in the Milky Way to distant galaxies at low and high redshift, including starburst galaxies. Extragalactic studies focus primarily on Kennicutt-Schmidt-type relations, i.e., measuring the SFR column density, $\sigsfr$, as a function of gas column density, $\siggas$ \citep[e.g.,][]{1998ApJ...498..541K,2008AJ....136.2846B,2012ARA&A..50..531K}, like in Figure~\ref{fig:sfrcoldens}.

In a recent theoretical effort with comparison to observations, \citet{2012ApJ...745...69K} suggested a unification of star formation in the Milky Way and in distant galaxies at low and high redshift into a universal law where $\sigsfr$ is about 1\% of the gas collapse rate, $\siggas/\tff$. This model is purely empirical, but fits observations with $\sigsfr$ varying over five orders of magnitude quite well. Yet, a physical foundation of this empirical model is still missing; it must of course ultimately be a result of the physical laws governing star formation that we have discussed in this review, with the apparent uniformity a result of a corresponding uniformity of the particular combination of physical properties that determine the SFR.  The remaining scatter in $\sigsfr$ around the proposed universal law is likely to reflect, in addition to an unavoidable observational scatter, local statistical fluctuations in physical properties.

Adding more observational data, including all local cloud and YSO measurements from \citet{2010ApJ...723.1019H}, \citet{2011ApJ...739...84G}, the Central Molecular Zone \citep{2009ApJ...702..178Y}, and the Small Magellanic Cloud \citep{2011ApJ...741...12B} in addition to the low- and high-redshift disk and starburst galaxies, \citet{2013arXiv1307.1467F} confirms the relatively tight correlation of $\sigsfr$ with $\siggas/\tff$ (compared to $\siggas$ only) and also suggests a likely contribution to the scatter in this correlation. \citet{2013arXiv1307.1467F} included simulation data with different forcing of the turbulence, sonic Mach numbers ranging from 5--50, and varying magnetic field strength in the \citet{2012ApJ...745...69K} framework, showing that a significant fraction of the scatter might be explained by variations in the properties of the turbulence, in particular by variations in the sonic Mach number of the star-forming clouds. Applying the theoretical models for the local SFR derived and discussed in this chapter, \citet{2013arXiv1307.1467F} shows that they are consistent with the simulation data and with the notion that the scatter seen in the observations can be explained by variations in the parameters of the turbulence. To test this suggestion, it will be necessary to measure the parameters of the turbulence, in particular the virial parameter, the sonic and Alfv\'en Mach numbers, the driving parameter $b$, and the star formation efficiency of the star-forming portions of all the objects contributing to the $\sigsfr$--$\siggas/\tff$ relation.

Specific tasks for future studies include:

1) Clarify definitions of MCs, both from the point of view of observations alone, and, more importantly, for establishing a better basis for comparing simulations with observations. This will become easier when using larger-scale simulations, where clouds can be selected using essentially the same methods as in observations.

2) Investigate the dense-gas mass fraction of star-forming clouds as a direct test of theoretical predictions (irrespective of a determination of the SFR). This could include a detailed numerical and observational study of the relation between the critical density of SFR theories, and the density where the PDF departs from the log-normal. 

3) Compare numerical simulations with observations avoiding the (necessarily very uncertain) `inversion' of observations into fundamental parameters, and focusing instead on the use of statistical `fingerprints' that can be readily constructed from observations and simulations alike. These may include spatial correlation functions of cores or YSOs, as well as column density PDFs.

4) Determine how to relate the SFR from counting YSOs in local clouds to indirect methods for more distant clouds and to global averages, using extragalactic methods. In such an effort, which aims at `calibrating' the extragalactic methods, the use of `forward analysis' based on numerical simulations (constructing synthetic observations from simulations) may constitute a useful tool.

5) Extend studies of the SFR versus gas properties to regions forming massive stars and clusters. These typically are denser and more turbulent and may reveal the importance of parameters like Mach number and turbulent forcing. Because they are more distant, different techniques are needed to study both the cloud structure and the star formation properties. From the modeling perspective they also pose challenges, because chemistry and radiative energy transfer is needed for a quantitative comparison with observations.

6) Explore even larger scales, for greater variety of SF conditions, to avoid modeling star-forming clouds as isolated systems with artificial boundary and initial conditions, and to study the role of large-scale feedbacks and driving. This will be achieved with higher-resolution extragalactic observations adequately interpreted \citep{2013MNRAS_K}, and with multi-scale star-formation simulations (whole galaxies or galactic fountains) that account for realistic driving forces such as galactic dynamics, SN explosions, stellar radiation and stellar outflows \citep{2004RvMP...76..125M}. 

Ultimately, the point of view taken in most of this review (as well as in the research papers reviewed), namely to explore and understand how the SFR depends on fundamental physical properties of the ISM, needs to be combined with the complementary view: asking how these fundamental properties of the ISM are, in turn, affected by star formation, and to what extent they are also modulated and affected by other external factors, such as galactic dynamics and density waves.

Given that we already know, from numerical simulations and theoretical modeling consistent with the simulations, that the SFR is a very steep function of the virial parameter, in particular at SFR values as low as those that are observed, one conclusion is practically unavoidable:  there must be a strong feedback from the average SFR level back to the physical properties that, in turn, determine the SFR. If one assumed, hypothetically, that the SFR was entirely controlled by external factors, with for example the normalization of the turbulent cascade (the constant factor in Larson's velocity relation) being entirely determined by galactic dynamics, with no influence of the feedback from star formation, that would be tantamount to claiming that the rather universal observed $\sfrff$ of $\sim1\%$ was a coincidental outcome of these external agents, maintaining a sufficiently uniform virial parameter for the SFR to come out right (or else manipulating also other variables to serendipitously compensate for virial parameter variations).  This appears much less likely than to assume that the feedback from stars closes a `servo loop', which keeps the SFR at a value that corresponds to consistency between the integrated feedback (SN explosions, outflows, and radiation) generated by star formation, and the distribution of physical parameters (particularly the virial parameter) that these feedbacks can sustain. 

External factors such as galactic dynamics are surely present, but star formation must play an important role for this servo loop to be effective. A possible problem with this idea is that the feedback is much stronger and different in regions forming massive stars, yet the star formation rate per free-fall time is similar. However, the existence of a pervasive inertial range in velocity dispersion, covering a large range of scales, shows that, whatever the ultimate source of the driving, the mechanical energy is distributed rather efficiently in space, being thus able to power also low mass star formation. The relative uniformity of the SFR may be just another consequence of this. 
Determining the ratio of contributions from stellar feedback relative to the total driving, and the resulting quasi-stable state of the feedback loop, via a combination of numerical and theoretical modeling, with the eventual success gauged by the consistency with observations, must indeed be considered as the ultimate goal of research into star formation and the SFR. When the fundamental properties of the ISM, such as the (distribution of) virial parameters, Mach numbers, and driving modes come out self-consistently from such modeling, with essentially no free parameters, then that goal has been reached.

\vspace{1cm}

\textbf{Acknowledgments.}
We thank Ralf Klessen, the anonymous referee, Patrick Hennebelle, Alexei Kritsuk, and Mark Krumholz for reading the 
manuscript and providing useful comments.
PP is supported by the FP7-PEOPLE-2010-RG grant PIRG07-GA-2010- 261359. Simulations by PP were carried out on the NASA/Ames Pleiades 
supercomputer, and under the PRACE project pra50751 running on SuperMUC at the LRZ (project ID pr86li).
CF thanks for support from the Australian Research Council for a Discovery Projects Fellowship (Grant DP110102191). NJE was supported by NSF 
Grant AST-1109116 to the University of Texas at Austin. The research of CFM is supported in part by NSF grant AST-1211729 and NASA grant 
NNX13AB84G. DJ  is supported by the National Research Council of Canada and by a Natural Sciences and Engineering Research Council of 
Canada (NSERC) Discovery Grant. JKJ is supported by a Lundbeck Foundation Junior Group Leader Fellowship. Research at Centre for Star and 
Planet Formation was funded by the Danish National Research Foundation and the University of CopenhagenÕs Programme of Excellence.  
Supercomputing time at  Leibniz Rechenzentrum (PRACE projects pr86li, pr89mu, and project pr32lo), at Forschungszentrum J{\"u}lich (project hhd20), 
and at DeIC/KU in Copenhagen are gratefully acknowledged.

\bigskip

\bibliographystyle{ppvi_lim1}

\end{document}